\documentclass[aps,prb]{revtex4}
\usepackage{amsmath,amssymb}
\usepackage{graphicx}
\usepackage{dcolumn}
\usepackage{bm}
\usepackage{color}
\usepackage{soul}
\DeclareMathAlphabet{\mathbbmsl}{U}{bbm}{m}{sl}
\begin{document}

\title{Self-consistent quantum-kinetic theory for interplay between pulsed-laser excitation and nonlinear carrier transport in a quantum-wire array}
	
\author{Jeremy R. Gulley}
\email[]{jgulley@kennesaw.edu}
\affiliation{Department of Physics, Kennesaw State University, Kennesaw, Georgia 30144, USA}
	
\author{Danhong Huang}
\affiliation{Air Force Research Laboratory, Space Vehicles Directorate, Kirtland Air Force Base, New Mexico 87117, USA}
\date{February 28, 2019}
	
\begin{abstract}
We propose a self-consistent many-body theory for coupling the ultrafast dipole-transition and carrier-plasma dynamics in a linear array of quantum wires with the scattering and absorption of ultrashort laser pulses.
The quantum-wire non-thermal carrier occupations are further driven by an applied DC electric field along the wires in the presence of resistive forces from intrinsic phonon and Coulomb scattering of photo-excited carriers.
The same strong DC field greatly modifies the non-equilibrium properties of the induced electron-hole plasma coupled to the propagating light pulse, while the induced longitudinal polarization fields of each wire significantly alters the nonlocal 
optical response from neighboring wires.
Here, we clarify several fundamental physics issues in this laser-coupled quantum wire system, including
laser influence on local transient photo-currents, photoluminescence spectra, and the effect of nonlinear transport in a micro-scale system on laser pulse propagation.
Meanwhile, we also anticipate some applications from this work, such as specifying the best combination of pulse sequence through a quantum-wire array to generate a desired THz spectrum 
and applying ultra-fast optical modulations to nonlinear carrier transport in nanowires.
\end{abstract}
\maketitle

\section{Introduction}
\label{sec1}

Several studies on strong light-matter interactions in semiconductors\,\cite{hole-burning,od,BuschlingerPRB2015} were reported in the past three decades. However, most of these studies involved some
non-self-consistent phenomenological models with model-parameter inputs from experimental observations. For example, a very early work on spectral-hole burning in the gain spectrum in Ref.\,[\onlinecite{hole-burning}]
assumed a constant pumping electric field by fully neglecting the field dynamics but including the electron dynamics instead under the energy-relaxation approximation. Later, such a simplified study was improved
by including full quantum kinetics for collisions between pairs of electrons in Ref.\,[\onlinecite{od}] within the second-order Born approximation. However, a spatially-uniform electric field was still adopted in their
model and the dynamics of this pumping electric field was not taken into account. Only very recently, a self-consistent calculation based on coupled Maxwell-Bloch equations was carried outin Ref.\,[\onlinecite{BuschlingerPRB2015}] for a multi-subband quantum-wire system. However, some phenomenological parameters were introduced for optical-coherence dephasing, spontaneous-emission rate
and energy-relaxation rate. Although the field dynamics was solved using Maxwell's equations in Ref.\,[\onlinecite{BuschlingerPRB2015}], only the propagating transverse electromagnetic field was studied while the localized longitudinal electromagnetic field
was excluded. It is interesting to point out that none of this early work on strong light-matter interactions has ever considered the drifting effects of electrons under a bias voltage in a self-consistent way.
\medskip

For the first time, we have established a unified quantum-kinetic model for both optical transitions and nonlinear transport of electrons within a single frame.
This unified quantum-kinetic theory is further coupled self-consistently to Maxwell's equations for the field propagation so as to study strong interactions between an ultrafast light pulse and driven electrons within a linear array of quantum wires
beyond the perturbation approach.
In our theory, the optical excitations of quantum-wire electrons by both propagating transverse and localized longitudinal electric fields are considered.
Meanwhile, the back action of optical polarizations\,\cite{backaction}, resulting from induced dynamical dipole moments and plasma waves due to electron-density fluctuations, on transverse and longitudinal electric fields is also included.
Moreover, the semiconductor Bloch equations\,\cite{od} (SBEs) are generalized to account for possible crystal momentum altering (non-vertical) transitions of electrons under a spatially nonuniform optical field, as well as the drifting of electrons under a net driving force including electron momentum dissipation.
\medskip

It is well known that photons do not interact directly with themselves. Instead, they interact indirectly through exciting electrons in nonlinear materials.
Although the strong interaction of photons in a laser pulse with electrons in quantum wires is extremely short in time and confined only within a micro-scale, photons still acquire ``fingerprints'' from the configuration space of excited electron-hole pairs.
These pairs can be detected through either a delayed light pulse in the same direction or another light beam in different directions as a stored photon quantum memory\,\cite{op-memory} from the first light pulse.
Within the perturbation regime, the nonlinear optical response of incident light can be studied, such as the Kerr effect and sum-frequency generation.\,\cite{nlo}
On the other hand, for optical reading, writing, and memory these ultrafast processes cannot be fully described by perturbation theories since they involve an ultrafast and strong interaction between a light pulse and the material.
\medskip

From the physics perspective, in this work we want to focus on three fundamental issues for a pulsed-laser irradiated quantum wire system. 
They are: variations in local transient photo-current and photoluminescence spectra by an incident laser pulse,
changes in propagation of laser pulses by nonlinear photo-carrier transport in a micro-scale quantum-wire array, and
optical reading of photon quantum memory (or electronic-excitation configurations) stored in a micro-scale quantum-wire array by a laser pulse.
From the technology perspective, however, we look for a specification of the best combination of pulse sequence through a quantum-wire array to generate a desired terahertz spectrum 
and a realization of ultra-fast optical control of nonlinear carrier transport in wires by laser pulses.
\medskip

The rest of this paper is organized as follows. In Sec.\,\ref{sec2}, we first establish a self-consistent formalism for propagation of laser pulses and generation of local optical-polarization fields by photo-excited electron-hole pairs in quantum wires.
After this, we develop in Sec.\,\ref{sec:SBE} another self-consistent theory for pulsed-laser excitation of electron-hole pairs and nonlinear transport of photo-excited carriers under a DC electric field.
Meanwhile, we also derive dynamical equations in Sec.\,\ref{interac} for describing back actions of electrons in quantum wires on interacting laser photons. 
In Sec.\,\ref{sandm}, we present a discussion of numerical results for transient properties of photo-excited carriers and laser pulses as well as for light-wire interaction dynamics.
Finally, conclusions are given in Sec.\,\ref{conclu} with some remarks.

\section{Pulse Propagation}
\label{sec2}

The pulse propagation is governed by Maxwell's equations which we solve using a Psuedo-Spectral Time Domain (PSTD) method\,\cite{taflove}.  Under this scheme, derivatives in real position (\mbox{\boldmath$r$}) space are evaluated in the Fourier wavevector (\mbox{\boldmath$q$}) space, in which Maxwell's equations take the form
\begin{subequations}
\begin{align}
 i \mbox{\boldmath$q$} \cdot \tilde{\mbox{\boldmath$D$}}(\mbox{\boldmath$q$},t) &= \tilde{\rho}_{\rm qw}(\mbox{\boldmath$q$},t)\ ,
\label{eq:ME1}
\\
 i \mbox{\boldmath$q$} \cdot \tilde{\mbox{\boldmath$B$}}(\mbox{\boldmath$q$},t) &= 0\ ,
\label{eq:ME2}
\\
 i \mbox{\boldmath$q$} \times \tilde{\mbox{\boldmath$E$}}(\mbox{\boldmath$q$},t) &= - \frac{\partial}{\partial t}  \tilde{\mbox{\boldmath$B$}}(\mbox{\boldmath$q$},t)\ ,
\label{eq:ME3}
\\
 i \mbox{\boldmath$q$} \times \tilde{\mbox{\boldmath$H$}}(\mbox{\boldmath$q$},t) &= \frac{\partial}{\partial t} \tilde{\mbox{\boldmath$D$}}(\mbox{\boldmath$q$},t)\ .
\label{eq:ME4}
\end{align}
\end{subequations}
Here, $\tilde{\mbox{\boldmath$E$}}(\mbox{\boldmath$q$},t)$ and $\tilde{\mbox{\boldmath$B$}}(\mbox{\boldmath$q$},t)$ represent the electric and magnetic fields, $\tilde{\mbox{\boldmath$D$}}(\mbox{\boldmath$q$},t)$ and $\tilde{\mbox{\boldmath$H$}}(\mbox{\boldmath$q$},t)$ are the auxiliary electric and magnetic fields,
and $\tilde{\rho}_{\rm qw}(\mbox{\boldmath$q$},t)$ is the charge-density distribution in the quantum wires embedded within a dielectric host. The two-dimensional (2D) Fourier transforms with respect to spatial positions are defined by
\begin{subequations}
\begin{align}
\tilde{f}(\mbox{\boldmath$q$}) & = \int d^2\mbox{\boldmath$r$} \,\texttt{e}^{-i{\bf q} \cdot {\bf r} }\,f(\mbox{\boldmath$r$})\ ,
\\
f(\mbox{\boldmath$r$}) & =\frac{1}{(2\pi)^2}\int d^2\mbox{\boldmath$q$} \,\texttt{e}^{i{\bf q} \cdot{\bf r}}\,\tilde{f}(\mbox{\boldmath$q$})\ .
\end{align}
\end{subequations}
In this work for non-magnetic materials, we neglect magnetic effects on the propagation and on the quantum wires so that the auxiliary magnetic field $\tilde{\mbox{\boldmath$H$}}(\mbox{\boldmath$q$},t) = \tilde{\mbox{\boldmath$B$}}(\mbox{\boldmath$q$},t)/ \mu_0$
with $\mu_0$ as the vacuum permeability.
We further divide the fields $(\tilde{\mbox{\boldmath$f$}})$ into transverse $(\tilde{\mbox{\boldmath$f$}}^\perp)$ and longitudinal $(\tilde{\mbox{\boldmath$f$}}^\|)$ contributions with respect to $\mbox{\boldmath$q$}$, defined by $\mbox{\boldmath$q$} \cdot\tilde{\mbox{\boldmath$f$}}^\perp = 0$ and $\mbox{\boldmath$q$} \times \tilde{\mbox{\boldmath$f$}}^\| = 0$, respectively.
By definition and from Eqs.\,\eqref{eq:ME1} and \eqref{eq:ME2} then, the longitudinal components of the auxiliary fields are given at all times by
\begin{subequations}
\begin{align}
\label{lcurr}
\tilde{\mbox{\boldmath$D$}}^\|(\mbox{\boldmath$q$},t) & = \mbox{\boldmath$\hat{e}$}_{\bf q}\left[\frac{\tilde{\rho}_{\rm qw}(\mbox{\boldmath$q$},t)}{iq}\right]\ ,
\\
\tilde{\mbox{\boldmath$H$}}^\|(\mbox{\boldmath$q$},t) & = 0\ ,
\end{align}
\end{subequations}
$\mbox{\boldmath$\hat{e}$}_{\bf q}=\mbox{\boldmath$q$}/q$ is a unit vector specifying the $\mbox{\boldmath$q$}$ direction,
$\tilde{\mbox{\boldmath$D$}}^\|(\mbox{\boldmath$q$},t)$ includes the longitudinal polarization fields, $\tilde{\mbox{\boldmath$P$}}^\|_{\rm qw}(\mbox{\boldmath$q$},t)$,
of quantum wires, and the longitudinal-optical conductivity, $\tilde{\sigma}_{\rm op}^\|(\mbox{\boldmath$q$},t)$, is determined from the equation:\,\cite{gerhardts}
$\tilde{\sigma}_{\rm op}^\|(\mbox{\boldmath$q$},t)\tilde{\mbox{\boldmath$E$}}^\|(\mbox{\boldmath$q$},t)=\tilde{\mbox{\boldmath$J$}}_{\rm qw}^\|(\mbox{\boldmath$q$},t)\equiv\partial\tilde{\mbox{\boldmath$P$}}^\|_{\rm qw}(\mbox{\boldmath$q$},t)/\partial t$.
\medskip

We further recast the electric field $\tilde{\mbox{\boldmath$E$}}(\mbox{\boldmath$q$},t)$ in terms of the electric displacement $\tilde{\mbox{\boldmath$D$}}(\mbox{\boldmath$q$},t)$, the polarization fields of the host material, $\tilde{\mbox{\boldmath$P$}}_{\rm host}(\mbox{\boldmath$q$},t)$, and the quantum wires, $\tilde{\mbox{\boldmath$P$}}_{\rm qw}(\mbox{\boldmath$q$},t)$.
The dispersion in the host material will be important for ultrashort pulses.
We therefore use a frequency ($\omega$) dependent dielectric function for the host,
$\epsilon_{\rm host}(\mbox{\boldmath$r$}, \omega) = \epsilon_b + \sum\limits_i \chi_i(\mbox{\boldmath$r$}, \omega)$,
where $\epsilon_b$ is a static and uniform background constant and $\chi_i(\mbox{\boldmath$r$},\omega)$ is the polarizability of the $i$th local Lorentz oscillator for bound electrons such that
${\mbox{\boldmath$P$}}_i(\mbox{\boldmath$r$},\omega) = \epsilon_0\chi_i(\mbox{\boldmath$r$}, \omega)\,\mbox{\boldmath$E$}(\mbox{\boldmath$r$},\omega)$
with $\epsilon_0$ as the vacuum permittivity, where
${\mbox{\boldmath$E$}}(\mbox{\boldmath$r$},\omega)={\mbox{\boldmath$E$}}^\|(\mbox{\boldmath$r$},\omega)+\sum\limits_n{\mbox{\boldmath$E$}}_n^\perp(\mbox{\boldmath$r$},\omega)$.
By solving a time-domain auxiliary differential equation for each $i${\rm th} oscillator,\,\cite{taflove} we get $\tilde{\mbox{\boldmath$P$}}_{\rm host}(\mbox{\boldmath$q$},t) =\epsilon_0(\epsilon_b-1)\tilde{\mbox{\boldmath$E$}}(\mbox{\boldmath$q$},t)+\sum\limits_i\tilde{\mbox{\boldmath$P$}}_{i}(\mbox{\boldmath$q$},t)$.
\medskip

Therefore, the time-evolution of the transverse auxiliary fields for different light pulses can be obtained from Eqs.\,\eqref{eq:ME3} and \eqref{eq:ME4}:
\begin{subequations}
\begin{align}
\frac{\partial \tilde{\mbox{\boldmath$D$}}^\perp(\mbox{\boldmath$q$},t)}{\partial t}
& =
i \mbox{\boldmath$q$}\times \tilde{\mbox{\boldmath$H$}}^\perp(\mbox{\boldmath$q$},t)\ ,
\label{eq:dDdt}
\\
\frac{\partial \tilde{\mbox{\boldmath$H$}}^\perp(\mbox{\boldmath$q$},t)}{\partial t}
& =
- i\epsilon_0c^2\,  \mbox{\boldmath$q$}  \times \tilde{\mbox{\boldmath$E$}}^\perp(\mbox{\boldmath$q$},t)\ ,
\label{eq:dHdt}
\end{align}
\end{subequations}
where $c=(\mu_0\epsilon_0)^{-1/2}$ is the vacuum speed of light, $\tilde{\mbox{\boldmath$D$}}^\perp(\mbox{\boldmath$q$},t)$ includes the transverse polarization fields, $\tilde{\mbox{\boldmath$P$}}^\perp_{\rm qw}(\mbox{\boldmath$q$},t)$, of quantum wires,
and the transverse-optical conductivity $\tilde{\sigma}_{{\rm op}}^\perp(\mbox{\boldmath$q$},t)$ can be determined from\,\cite{gerhardts}
$\tilde{\sigma}_{{\rm op}}^\perp(\mbox{\boldmath$q$},t)\,\tilde{\mbox{\boldmath$E$}}^\perp(\mbox{\boldmath$q$},t)
=
\tilde{\mbox{\boldmath$J$}}_{\rm qw}^\perp(\mbox{\boldmath$q$},t)\equiv\partial \tilde{\mbox{\boldmath$P$}}^\perp_{\rm qw}(\mbox{\boldmath$q$},t)/\partial t$.
%Note that only the transverse components of $\tilde{\mbox{\boldmath$E$}}(\mbox{\boldmath$q$},t)$ are selected by the cross product in Eq.\,\eqref{eq:dHdt}.
At all times, the longitudinal ($\|$) and transverse ($\perp$) components of $\tilde{\mbox{\boldmath$E$}}(\mbox{\boldmath$q$},t)$ are evaluated through\,\cite{jackson}
%\begin{subequations}
\begin{align}
\label{eq:E}
\tilde{\mbox{\boldmath$E$}}^{\perp,\|}(\mbox{\boldmath$q$},t)
=
\frac{\tilde{\mbox{\boldmath$D$}}^{\perp,\|} (\mbox{\boldmath$q$},t)-\sum\limits_{i}\tilde{\mbox{\boldmath$P$}}^{\perp,\|}_{i}(\mbox{\boldmath$q$},t)+\tilde{\mbox{\boldmath$P$}}^{\perp,\|}_{\rm qw}(\mbox{\boldmath$q$},t )}{\epsilon_0\epsilon_b}\ .
\end{align}
%\end{subequations}
Note that $\tilde{\mbox{\boldmath$P$}}^{\{\|,\perp\}}_{\rm qw}(\mbox{\boldmath$q$},t )$ in Eq.\,\eqref{eq:E} has often been omitted.
Instead, it enters directly into Eq.\,\eqref{eq:ME4} as a term
$\tilde{\mbox{\boldmath$J$}}_{\rm qw}(\mbox{\boldmath$q$},t)=\partial \tilde{\mbox{\boldmath$P$}}_{\rm qw}(\mbox{\boldmath$q$},t)/\partial t$, mainly flowing along the quantum-wire direction $\mbox{\boldmath$\hat{e}$}_{\rm w}$ in a 2D field system.
\medskip

We orient all wires along the $\mbox{\boldmath$\hat{e}$}_{\rm w}$ direction, and split $\mbox{\boldmath$q$}$ into vectors parallel $\mbox{\boldmath$q$}_\| = (\mbox{\boldmath$\hat{e}$}_{\rm w}\cdot \mbox{\boldmath$q$})\,\mbox{\boldmath$\hat{e}$}_{\rm w}$ and perpendicular $\mbox{\boldmath$q$}_{\perp} = - (\mbox{\boldmath$\hat{e}$}_{\rm w} \times \mbox{\boldmath$\hat{e}$}_{\rm w}\times \mbox{\boldmath$q$})$ to $\mbox{\boldmath$\hat{e}$}_{\rm w}$.
Note that the directions of $\mbox{\boldmath$q$}_\|$ and $\mbox{\boldmath$q$}_\perp$ are not related to longitudinal and transverse contributions of an electromagnetic field. The quantum-wire source terms in Maxwell's equations are the sum of the contributions from each quantum wire and are expressed as\,\cite{huang-3}
\begin{subequations}
\begin{align}
\label{eq:rhoq}
\tilde{\rho}_{\rm qw} (\mbox{\boldmath$q$},t) & =
\sum_j  \tilde{\rho}^{\rm 1D}_{j} ({q}_\|,t) \;
\texttt{e}^{-i {\bf q_\perp} \cdot {\bf R}_j^\perp  \,- \, q^2_\perp / 4\alpha^2 }\ ,
\\
\label{eq:Pq1}
\tilde{\mbox{\boldmath$P$}}^{\{\|,\perp\}}_{\rm qw}(\mbox{\boldmath$q$},t) & =\sum_{\sigma=x,y}\tilde{\cal P}^\sigma_{\rm qw}(\mbox{\boldmath$q$},t)\,\tilde{\mbox{\boldmath${\cal G}$}}^\sigma_{\{\|,\perp\}}(\mbox{\boldmath$q$})=
\sum_j\,\texttt{e}^{-i {\bf q_\perp} \cdot {\bf R}^\perp_j  \,- \, q^2_\perp / 4\alpha^2 } \sum_{\sigma=x,y}\tilde{{P}}^\sigma_{j}(q_\|,t)\,\tilde{\mbox{\boldmath${\cal G}$}}^\sigma_{\{\|,\perp\}}(\mbox{\boldmath$q$})\ ,
\end{align}
\end{subequations}
where $\sigma=x,\,y$ label two of three independent dipole directions in a two-dimensional propagating system for electrons within a quantum wire,
the centered transverse position of the $j$th quantum wire in real space is denoted by $\mbox{\boldmath$R$}^\perp_j$, and the width of each wire is $2/\alpha$. The total electric field $\tilde{\mbox{\boldmath$E$}}(\mbox{\boldmath$q$},t)=\tilde{\mbox{\boldmath$E$}}^\|(\mbox{\boldmath$q$},t)
+\tilde{\mbox{\boldmath$E$}}^\perp(\mbox{\boldmath$q$},t)$ is a complex field and $|\tilde{\mbox{\boldmath$E$}}(\mbox{\boldmath$q$},t)|=\sqrt{|\tilde{{E}}^\perp(\mbox{\boldmath$q$},t)|^2+|\tilde{{E}}^\|(\mbox{\boldmath$q$},t)|^2}$.
In addition, we would like to emphasize that the quasi-one-dimensional (quasi-1D) quantum wire is still treated as a bulk semiconductor material for optical transitions of electrons.
The polarization field $\tilde{\mbox{\boldmath$P$}}_{\rm qw}(\mbox{\boldmath$q$},t)$ should point to the direction of 2D dipole moments.
For centrosymmetric GaAs cubic crystal with isotropic band structures at $\Gamma$-point, the unit vector in the dipole direction is
found to be $\mbox{\boldmath$\hat{e}$}^\sigma_{\bf d}=\mbox{\boldmath$\hat{e}$}_{x,y}$ with $\mbox{\boldmath$\hat{e}$}_{x,y}$ as two coordinate unit vectors.
The 1D field sources, $\tilde{\rho}^{\rm 1D}_{j} ({q}_\|,t)$ and $\tilde{{P}}^\sigma_{j}(q_\|,t)$, in Eqs.\,\eqref{eq:rhoq} and \eqref{eq:Pq1}
are calculated from the solutions to the SBEs in the 1D momentum space of the wire as described below. Moreover, $\tilde{\mbox{\boldmath${\cal G}$}}^\sigma_{\{\|,\perp\}}(\mbox{\boldmath$q$})$ in Eq.\,\eqref{eq:Pq1}
represent the two vector projection functions for longitudinal ($\|$) and transverse ($\perp$) directions of the polarization field, respectively. Specifically, we can write them down as\,\cite{huang-2}
\begin{subequations}
\begin{align}
\label{parapx}
\tilde{\mbox{\boldmath${\cal G}$}}^x_{\|}(\mbox{\boldmath$q$}) & =\left(\mbox{\boldmath$\hat{e}$}_{\bf q}\cdot\mbox{\boldmath$\hat{e}$}_x\right)\,\mbox{\boldmath$\hat{e}$}_{\bf q}
=\frac{q_\perp}{q_\perp^2+q_\|^2}\,(q_\perp\mbox{\boldmath$\hat{e}$}_x+q_\|\mbox{\boldmath$\hat{e}$}_y)\ ,
\\
\label{parapy}
\tilde{\mbox{\boldmath${\cal G}$}}^y_{\|}(\mbox{\boldmath$q$}) & =\left(\mbox{\boldmath$\hat{e}$}_{\bf q}\cdot\mbox{\boldmath$\hat{e}$}_y\right)\,\mbox{\boldmath$\hat{e}$}_{\bf q}
=\frac{q_\|}{q_\perp^2+q_\|^2}\,(q_\perp\mbox{\boldmath$\hat{e}$}_x+q_\|\mbox{\boldmath$\hat{e}$}_y)\ ,
\\
\label{perppx}
\tilde{\mbox{\boldmath${\cal G}$}}^x_{\perp}(\mbox{\boldmath$q$}) & =-\left(\mbox{\boldmath$\hat{e}$}_{\bf q}\times\mbox{\boldmath$\hat{e}$}_{\bf q}\times\mbox{\boldmath$\hat{e}$}_x\right)
=\frac{q_\|}{q_\perp^2+q_\|^2}\,(q_\|\mbox{\boldmath$\hat{e}$}_x-q_\perp\mbox{\boldmath$\hat{e}$}_y)\ ,
\\
\label{perppy}
\tilde{\mbox{\boldmath${\cal G}$}}^y_{\perp}(\mbox{\boldmath$q$}) & =-\left(\mbox{\boldmath$\hat{e}$}_{\bf q}\times\mbox{\boldmath$\hat{e}$}_{\bf q}\times\mbox{\boldmath$\hat{e}$}_y\right)
=\frac{-q_\perp}{q_\perp^2+q_\|^2}\,(q_\|\mbox{\boldmath$\hat{e}$}_x-q_\perp\mbox{\boldmath$\hat{e}$}_y)\ ,
\end{align}
\end{subequations}
where $\mbox{\boldmath$q$}=\{q_\perp\hat{\mbox{\boldmath$e$}}_x,\,q_\|\hat{\mbox{\boldmath$e$}}_y\}$ for our chosen
$\hat{\mbox{\boldmath$e$}}_{\rm w}=\hat{\mbox{\boldmath$e$}}_y$.

\section{Laser-Semiconductor Plasma Interaction}
\label{sec:SBE}

For photo-excited spin-degenerate electrons and holes in the $j$th quantum wire, the quantum-kinetic semiconductor Bloch equations are given by\,\cite{HaugKoch,mukamelSBEs,BuschlingerPRB2015}
\begin{subequations}
\begin{align}
\label{eq:dnedt}
\frac{dn^{\rm e}_{j, k}(t)}{dt}
= &
 \frac{2}{\hbar}  \sum_{k^\prime}\,  {\rm Im}  \left\{  \mbox{\boldmath$p$}_{j, k,k^\prime}(t) \cdot \mbox{\boldmath$\Omega$}_{j,k',k}(t) \right\}
+ \left.\frac{\partial n^{\rm e}_{j,k}(t)}{\partial t}\right|_{\rm rel}\ ,
\\
\label{eq:dnhdt}
\frac{dn^{\rm h}_{j, k^\prime}(t)}{dt}
= &
\frac{2}{\hbar}  \sum_{k}\, {\rm Im} \left\{\mbox{\boldmath$p$}_{j,k,k^\prime}(t) \cdot \mbox{\boldmath$\Omega$}_{j,k',k}(t) \right\}
+ \left.\frac{\partial n^{\rm h}_{j,k^\prime}(t)}{\partial t}\right|_{\rm rel}\ ,
\\
\label{eq:dpdt}
i \hbar\, \frac{d\mbox{\boldmath$p$}_{j,k,k^\prime}(t)}{dt}
& =\left[\varepsilon^{\rm e}_k  +   \varepsilon^{\rm h}_{k^\prime} +\varepsilon_{\rm G}  +  \Delta\varepsilon^{\rm e}_{j, k}+\Delta\varepsilon^{\rm h}_{j, k'}  - i \hbar\,\Delta^{\rm eh}_{j,k,k'}(t)\right]\mbox{\boldmath$p$}_{j,k,k^\prime}(t)
-\left[1 - n^{\rm e}_{k}(t) - n^{\rm h}_{k^\prime}(t)\right] \hbar \mbox{\boldmath$\Omega$}_{j,k,k^\prime}(t)\\
\nonumber
& +i\hbar\sum\limits_{q\neq 0}\,\Lambda^{\rm e}_{j,k,q}(t)\,\mbox{\boldmath$p$}_{j,k+q,k^\prime}(t)
+i\hbar\sum\limits_{q'\neq 0}\,\Lambda^{\rm h}_{j,k',q'}(t)\,\mbox{\boldmath$p$}_{j,k,k^\prime+q'}(t)\ ,
\end{align}
\end{subequations}
where $\mbox{\boldmath$p$}_{j,k,k^\prime}(t)=\sum\limits_{\sigma=x,y}\,p^\sigma_{j,k,k^\prime}(t)\,\mbox{\boldmath$\hat{e}$}_{\bf d}^{\sigma}$ are potentially
two equations with respect to $p^{x,y}_{j,k,k^\prime}(t)$ that are formally combined into one vector equation \eqref{eq:dpdt},
$\sigma=x,\,y$ correspond to the dipole directions,
the spin degeneracy of carriers is included, $\varepsilon_{\rm G}$ is the bandgap of a host semiconductor including size-quantization effects of quantum wires,
the retarded interwire electromagnetic coupling has been included in Eqs.\,\eqref{eq:E}, \eqref{eq:Pq1} and in Eqs.\,\eqref{eq:Rabix} and \eqref{eq:Rabiy}. In Eqs.\,\eqref{eq:dnedt}-\eqref{eq:dpdt},
$n^{\rm e}_{j,k}(t)$ and $n^{\rm h}_{j,k^\prime} (t)$ are the electron (e) and hole (h) occupation numbers at momenta $\hbar k$, and $\hbar k^\prime$, respectively, and $\hbar q$, $\hbar q'$ represent their transition momenta.
The quantum coherence between electron and hole states coupled to the electric field is $\mbox{\boldmath$p$}_{j,k,k^\prime}(t)$, $\mbox{\boldmath$\Omega$}_{j,k,k^\prime}(t)$ is the renormalized Rabi frequency,
$\varepsilon^{\rm e}_{j, k}$ and $\varepsilon^{\rm h}_{j, k'}$ indicate their kinetic energies,
and $\Delta\varepsilon^{\rm e}_{j, k}$ and $\Delta\varepsilon^{\rm h}_{j, k'}$ are the Coulomb renormalization\,\cite{huang-4} of the kinetic energies of electrons and holes.
Moreover, $\Delta^{\rm eh}_{j,k,k'}(t)=\Delta^{\rm e}_{j,k}(t)+\Delta^{\rm h}_{j,k'}(t)$ is the diagonal dephasing rate\,\cite{od} (quasi-particle lifetime),
while $\Lambda^{\rm e}_{j,k,q}(t)$ and $\Lambda^{\rm h}_{j,k',q'}(t)$ are
the off-diagonal dephasing rates\,\cite{od} (pair-scattering) for electrons and holes (see Appendix\ \ref{app-4} for details).
\medskip

In deriving the above equations, the electron and hole wave functions in a quantum wire are assumed to be
$\Psi^{\rm e,h}_{k}(\mbox{\boldmath$\xi$})=\psi^{\rm e,h}_0(\mbox{\boldmath$\xi$}_\perp)\, \exp(i k\xi_\|) / \sqrt{{\cal L}}$,
where ${\cal L}$ represents the length of a quantum wire,
$\psi^{\rm e,h}_0(\mbox{\boldmath$\xi$}_\perp)= (\alpha_{\rm e,h} / \sqrt{\pi})\, \exp \left( -\alpha_{\rm e,h}^2 \xi_\perp^2 / 2 \right)$
are the ground-state wavefunctions of electrons and holes in two transverse directions,
$\alpha_{\rm e,h}=\sqrt{m^\ast_{\rm e,h}\Omega_{\rm e,h}/\hbar}$, $m^\ast_{\rm e,h}$ are the electron and hole effective masses,
$\hbar\Omega_{\rm e,h}$ are the level separations between the ground and the first excited state of electrons and holes due to finite-size quantization, and $\alpha$ in Eqs.\,\eqref{eq:rhoq}-\eqref{eq:Pq1} is given
by $2/\alpha=1/\alpha_{\rm e}+1/\alpha_{\rm e}$, and the local position vector $\mbox{\boldmath$\xi$} = \{\mbox{\boldmath$\xi$}_\perp, \xi_\| \}$ just as $\mbox{\boldmath$q$} = \{\mbox{\boldmath$q$}_\perp, q_\| \}$ earlier.
The dipole-coupling matrix element is calculated as $d_{\rm cv}=\sqrt{(3e^2\hbar^2/4m_0\varepsilon_{\rm G})\,[(m_0/m_{\rm e}^\ast)-1]}$ for the isotropic interband dipole moment at the $\Gamma$-point\,\cite{huang-5} and $m_0$ is the free-electron mass.
If the quantum-kinetic occupations $n^{\rm e,h}_{j, k}(t)$ in Eqs.\,\eqref{eq:dnedt} and \eqref{eq:dnhdt} are replaced by their thermal-equilibrium Fermi functions $n_0(\varepsilon^{\rm e,h}_k)$ and the Rabi frequencies $\Omega^{x,y}_{j,k,k^\prime}(t)$ in Eq.\,\eqref{eq:dpdt} are also replaced by
$d_{\rm cv}E_{x,y}^{(0)}/\hbar$ for an incident electric field, we arrive at the optical linear-response theory from Eq.\,\eqref{eq:dpdt}
after neglecting all dephasing terms.
\medskip

In Eq.\,\eqref{eq:dpdt}, $\varepsilon^{\rm e}_k=\hbar^2 k^2/2m_{\rm e}^\ast$
and   $\varepsilon^{\rm h}_{k^\prime}=\hbar^2 k^{\prime\,2}/2m_{\rm h}^\ast$
are the kinetic energies of electrons and holes.
Their correction terms, $\Delta\varepsilon^{\rm e}_{j,k}$ and $\Delta\varepsilon^{\rm h}_{j,k'}$, are given by:\,\cite{huang-4}
\begin{subequations}
\begin{align}
\label{eq:DelEk}
\Delta\varepsilon^{\rm e}_{j, k}
& =2\sum_{q}\,n^{\rm e}_{j, q}(t)  V^{\rm ee}_{k,q;\,q,k}
-  \sum_{q \ne k}\,  n^{\rm e}_{j, q}(t)  V^{\rm ee}_{k,q;\,k,q}
-  2\sum_{q^\prime} \, n^{\rm h}_{j, q^\prime}(t) \,  V^{\rm eh}_{k,q^\prime ;\, q^\prime,k}\ ,
\\
\label{eq:DelEkp}
\Delta\varepsilon^{\rm h}_{j, k'}
& =2 \sum_{q^\prime}  \,  n^{\rm h}_{j, q^\prime}(t)   V^{\rm hh}_{k^\prime,q^\prime;\,q^\prime,k^\prime}
  -  \sum_{q^\prime \ne k'}  \, n^{\rm h}_{j, q^\prime}(t)V^{\rm hh}_{k^\prime,q^\prime;\,k^\prime,q^\prime}
-  2\sum_{q} \,             n^{\rm e}_{j, q}(t)            \,  V^{\rm eh}_{q,k^\prime ;\, k^\prime,q}\ ,
\end{align}
\end{subequations}
which also account for the excitonic interaction energy.
The Coulomb-interaction matrix elements, $V^{\rm eh}_{k_1,k_1^\prime;\,k_2^\prime,k_2}$, $V^{\rm hh}_{k^\prime_1,k^\prime_2;\,k^\prime_3,k^\prime_4}$ and
$V^{\rm ee}_{k_1,k_2;\,k_3,k_4}$, introduced in Eqs.\,\eqref{eq:DelEk}, \eqref{eq:DelEkp}, \eqref{eq:Rabix} and \eqref{eq:Rabiy} are explicitly given in Appendix\ \ref{app2}.
\medskip

In the presence of many photo-excited carriers, i.e., for the total numbers of electrons $N_{\rm e}(t)$ and holes $N_{\rm h}(t)$, the Coulomb interaction will be screened by a dielectric function $\epsilon_{\rm 1D}(q_\|,\,t)$ in the Thomas-Fermi limit\,\cite{huang-7},
{\em e.g.\/}, $V^{\rm ee}_{k_1,k_2;\,k_3,k_4}\to V^{\rm ee}_{k_1,k_2;\,k_3,k_4}/\epsilon_{\rm 1D}(|k_1-k_4|,\,t)$,
$V^{\rm hh}_{k^\prime_1,k^\prime_2;\,k^\prime_3,k^\prime_4}\to V^{\rm hh}_{k^\prime_1,k^\prime_2;\,k^\prime_3,k^\prime_4}/\epsilon_{\rm 1D}(|k^\prime_4-k^\prime_1|,\,t)$ and
$V^{\rm eh}_{k_1,k^\prime_1;\,k^\prime_2,k_2}\to V^{\rm eh}_{k_1,k^\prime_1;\,k^\prime_2,k_2}/\epsilon_{\rm 1D}(|k_1-k_2|,\,t)$.
Using the high-density random-phase approximation (RPA) at low temperatures, $\epsilon_{\rm 1D}(q_\|,\,t)$ is calculated as\,\cite{book-huang}
\begin{equation}
	\epsilon_{\rm 1D}(q_\|,\,t)
	=
	1 -
	\lim_{\omega\rightarrow 0} \,
	\frac{ 2 \beta m_{\rm e}^\ast}{\pi \hbar^2 q_\|}\,
	\ln\left\{
				\frac{\omega^2-[\Omega_{\rm e}^-(q_\|,t)]^2}{\omega^2-[\Omega^+_{\rm e}(q_\|,t)]^2}
		\right\}  \,
		K_0(q_\|R_e)
	-\lim_{\omega\rightarrow 0}  \,
	\frac{2 \beta m_{\rm h}^\ast}{\pi \hbar^2 q_\|}\,
	\ln\left\{
				\frac{\omega^2-[\Omega_{\rm h}^-(q_\|,t)]^2}{\omega^2-[\Omega^+_{\rm h}(q_\|,t)]^2}
	\right\}  \,
	K_0(q_\|R_h) \ ,
	\label{eq:E1D}
\end{equation}
where $q_\|$ is the absolute value of the electron wave number,
$\beta=e^2/4\pi\epsilon_0\epsilon_r$ with $\epsilon_r$ as the average dielectric constant of the quantum wire,
$K_0(q_\||x|)$ is the modified Bessel function of the third kind,
$\Omega_{\rm e,h}^{\pm}(q_\|,t)=(\hbar q_\|/2m_{\rm e,h}^\ast)\,|q_\|\pm 2k^{\rm e,h}_{\rm F}(t)|$,
$k^{\rm e,h}_{\rm F}(t)=\pi n_{\rm 1D}^{\rm e,h}(t)/2$
are the Fermi wavelengths,
$R_{\rm e,h}=\sqrt{(2/\alpha_{\rm e,h})^2+\delta_0^2}$, $\delta_0$ is the thickness of a quantum wire,
and
$n_{\rm 1D}^{\rm e,h}(t)=N_{\rm e,h}(t)/{\cal L}$
are the linear densities of photo-excited carriers.
\medskip

The additional relaxation terms in Eqs.\,\eqref{eq:dnedt} and \eqref{eq:dnhdt} are given by\,\cite{PhysRevB.69.075214}
\begin{subequations}
\begin{align}
\label{eq:dnedtREL}
\left.\frac{\partial n^{\rm e}_{j,k}(t)}{\partial t}\right|_{\rm rel}
& =\left.\frac{\partial n^{\rm e}_{j,k}(t)}{\partial t}\right|_{\rm scat}
- {\cal R}_{j,{\rm sp}}(k,\,t)   \,   n^{\rm e}_{j,k}(t)   \,   n^{\rm h}_{j,k}(t)
+ \frac{{\cal F}_j^{\rm e}(t)}{\hbar}\,\frac{\partial n^{\rm e}_{j,k}(t)}{\partial k}\ ,
\\
	\label{eq:dnhdtREL}
	\left.
	\frac{\partial n^{\rm h}_{j,k'}(t)}{\partial t}
	\right|_{\rm rel}
&	=
	\left.
	\frac{\partial n^{\rm h}_{j,k'}(t)}{\partial t}
	\right|_{\rm scat}
	- {\cal R}_{j,{\rm sp}}(k',\,t)  \,   n^{\rm e}_{j,k'}(t)   \,   n^{\rm h}_{j,k'}(t)
	- \frac{{\cal F}_j^{\rm h}(t)}{\hbar} \,\frac{\partial n^{\rm h}_{j, k^\prime} (t)}{\partial k^\prime}\ .
\end{align}
\end{subequations}
Here, on the right-hand side, the first term describes non-radiative energy relaxation through Coulomb and phonon scattering,
the second term corresponds to spontaneous recombinations of e-h pairs, and the last term represents carrier drifting in the presence of an applied DC electric field.
\medskip

The Boltzmann-type scattering terms for non-radiative energy relaxation in Eqs.\,\eqref{eq:dnedtREL} and \eqref{eq:dnhdtREL} are given by\,\cite{huang-6}
\begin{subequations}
\begin{align}
\left.\frac{\partial n^{\rm e}_{j,k}(t)}{\partial t}\right|_{\rm scat}
= &
W^{\rm e, (in)}_{j, k}(t)
\left[1 - n^{\rm e}_{j,k}(t)\right]
- W^{\rm e, (out)}_{j, k}(t)  \,  n^{\rm e}_{j,k}(t)\ ,
\label{eq:dnedtSCAT}
\\
\left.
\frac{\partial n^{\rm h}_{j,k'}(t)}{\partial t}
\right|_{\rm scat}
= &
W^{\rm h, (in)}_{j, k'}(t)
\left[1 - n^{\rm h}_{j,k'}(t)\right]
- W^{\rm h, (out)}_{j, k'}(t)  \,  n^{\rm h}_{j,k'}(t)\ ,
\label{eq:dnhdtSCAT}
\end{align}
\end{subequations}
where the explicit expressions for scattering-in, $W^{\rm e,h, (in)}_{j, k}(t)$, and scattering-out, $W^{\rm e,h, (out)}_{j, k}(t)$, rates for electrons and holes are
presented in Appendix\ \ref{app-3}.
\medskip

For hot photo-excited carriers in non-thermal occupations, the time-dependent spontaneous-emission rate,
${\cal R}_{j,{\rm sp}}(k,\,t)$, introduced in Eqs.\,\eqref{eq:dnedtREL} and \eqref{eq:dnhdtREL} for each quantum wire is calculated as\,\cite{huang-1}
\begin{align}
\nonumber
	\label{eq:Rsp}
		{\cal R}_{j,{\rm sp}}(k,\,t)& =
	\frac{3d_{\rm cv}^2}{\epsilon_0  \sqrt{\epsilon_{\rm r}}}\int_0^{\infty} d\omega^\prime  \,
	\Bigg\{
	\hbar \omega^\prime  \,
	\rho_0(\omega^\prime)\,
    L(\hbar\omega^\prime  	-  \varepsilon_{\rm G} - \varepsilon^{\rm e}_k
	-  \varepsilon^{\rm h}_k  -  \varepsilon_{j,{\rm c}}(k,\,t), \; \hbar\gamma_{\rm eh})\,
\\
	& \times
	    M(\hbar \omega' - \varepsilon_{\rm G} - \varepsilon_{j,{\rm c}}(k=0,\,t), \; \hbar\gamma_{\rm eh})
	\Bigg\}\ ,
\end{align}
where $L(a,b)=(b/\pi)/(a^2+b^2)$ is the Lorentzian line-shape function, $M(a,b)=[1+(2/\pi)\tan^{-1}(a/b)]/2$ is the broadened step function,
$\gamma_{\rm eh}=(\gamma_{\rm e}+\gamma_{\rm h})/2$ with $1/\gamma_{\rm eh}$ as the lifetime of photo-excited non-interacting electrons (holes),
and $\rho_0(\omega)=\omega^2/c^3\pi^2\hbar$ is the density-of-states for spontaneously-emitted photons in vacuum.
Moreover,  the Coulomb renormalization $\varepsilon_{j,{\rm c}}(k,\,t)$ of the transition energy
in the $j$th quantum wire is found to be
\begin{align}
\nonumber
	\varepsilon_{j,{\rm c}}(k,\,t)
	& =
	\sum_{q}  \,  n^{\rm e}_{j, q}(t)   \left(V^{\rm ee}_{k,q;\,q,k} - V^{\rm ee}_{k,q;\,k,q}\right)
	+\sum_{q^\prime}  \,n^{\rm h}_{j, q^\prime}(t)
	\left( V^{\rm hh}_{k,q^\prime;\,q^\prime,k}  -  V^{\rm hh}_{k,q^\prime;\,k,q^\prime}  \right)
	\\ &
\label{eq:Ec}
	-\sum_{q\neq k}  \,  n^{\rm e}_{j, q}(t)  \,  V^{\rm eh}_{q,k; \, k,q}
	-\sum_{q^\prime\neq k}  \,  n^{\rm h}_{j, q^\prime}(t)  \,  V^{\rm eh}_{k,q^\prime; \,q^\prime,k}
	-V^{\rm eh}_{k,k;\,k,k}\ ,
\end{align}
where the first two terms are associated with the Hartree-Fock energies\,\cite{huang-7} for electrons and holes, while the remaining terms are related to the excitonic interaction energy.
\medskip

The net driving forces, ${\cal F}_{j,{\rm e}}(t)$ and ${\cal F}_{j,{\rm h}}(t)$, introduced in Eqs.\eqref{eq:dnedtREL} and \,\eqref{eq:dnhdtREL} for electrons and  holes,
including the resistive ones from the optical-phonon scattering of photo-excited carriers, can be calculated from\,\cite{PhysRevB.71.195205}
\begin{subequations}
\begin{align}
	\label{eq:Fe}
	{\cal F}_j^{\rm e}(t)
&	=
	-eE_{\rm dc}
	-2\,\sum_{k,q} \,
		\hbar q
		\Bigg\{\Theta^{\rm em}_{j, \rm e}(k,q,\,t) -\Theta^{\rm abs}_{j, \rm e}(k,q,\,t)\Bigg\}\ ,
\\
	\label{eq:Fh}
	{\cal F}_j^{\rm h}(t)
&	=
	+eE_{\rm dc}
	-2\,\sum_{k',q'} \,
		\hbar q'
		\Bigg\{\Theta^{\rm em}_{j, \rm h}(k',q',\,t)-\Theta^{\rm abs}_{j, \rm h}(k',q',\,t)\Bigg\}\ ,
\end{align}
\end{subequations}
where $E_{\rm dc}$ is the applied DC electric field. In Eqs.\,\eqref{eq:Fe} and \eqref{eq:Fh}, the emission (em) and absorption (abs) rates for longitudinal-optical phonons in intrinsic
and defect-free quantum wires are given by\,\cite{PhysRevB.71.195205}
\begin{subequations}
\begin{align}
	\nonumber
	\Theta^{\rm em}_{j, \rm e}(k,q,\,t)
	=
	&
	\frac{4\pi}{\hbar}  \,
	\left|V^{\rm ep}_{k,k-q}\right|^2\,n_{j,k}^{\rm e}(t)\left[1-n_{j,k-q}^{\rm e}(t)\right]\left[N_0(\Omega_{\rm ph})+1\right]
	\\
	&
	\times
	L(\varepsilon^{\rm e}_{k-q}-\varepsilon^{\rm e}_{k} + \hbar\Omega_{\rm ph}-\hbar q\,v_j^{\rm e}(t),\gamma_{\rm e})\,
	\theta(\hbar\Omega_{\rm ph}-\hbar q\,v_j^{\rm e}(t))\ ,
	\label{eq:eem}
	\\
	\nonumber
	\Theta^{\rm abs}_{j, \rm e}(k,q,\,t)
	=
	&
	\frac{4\pi}{\hbar}  \,
	\left|V^{\rm ep}_{k,k-q}\right|^2\,n_{j,k-q}^{\rm e}(t)\left[1-n_{j,k}^{\rm e}(t)\right]N_0(\Omega_{\rm ph})
	\\
	&
	\times
	L(\varepsilon^{\rm e}_k-\varepsilon^{\rm e}_{k-q}-\hbar\Omega_{\rm ph}+\hbar q\,v_j^{\rm e}(t),\gamma_{\rm e})\,
		\theta(\hbar\Omega_{\rm ph}-\hbar q\,v_j^{\rm e}(t))\ ,
	\label{eq:eabs}
\end{align}
\end{subequations}
\begin{subequations}
	\begin{align}
	\nonumber
	\Theta^{\rm em}_{j, \rm h}(k',q',\,t)
	=
	&
	\frac{4\pi}{\hbar}  \,
	\left|V^{\rm hp}_{k',k'-q'}\right|^2\,n_{j,k'}^{\rm h}(t)\left[1-n_{j,k'-q'}^{\rm h}(t)\right]\left[N_0(\Omega_{\rm ph})+1\right]
	\\
	&
	\times
	L(\varepsilon^{\rm h}_{k'-q'}-\varepsilon^{\rm h}_{k'} + \hbar\Omega_{\rm ph}-\hbar q'v_j^{\rm h}(t),\gamma_{\rm h})\,
	\theta(\hbar\Omega_{\rm ph}-\hbar q'v_j^{\rm h}(t))\ ,
	\label{eq:hem}
	\\
	\nonumber
	\Theta^{\rm abs}_{j, \rm h}(k',q',\,t)
	=
	&
	\frac{4\pi}{\hbar}  \,
	\left|V^{\rm hp}_{k',k'-q'}\right|^2\,n_{j,k'-q'}^{\rm h}(t)\left[1-n_{j,k'}^{\rm h}(t)\right]N_0(\Omega_{\rm ph})
	\\
	&
	\times
	L(\varepsilon^{\rm h}_{k'}-\varepsilon^{\rm h}_{k'-q'}-\hbar\Omega_{\rm ph}+\hbar q'v_j^{\rm h}(t),\gamma_{\rm h})\,
	\theta(\hbar\Omega_{\rm ph}-\hbar q'v_j^{\rm h}(t))\ ,
	\label{eq:habs}
	\end{align}
\end{subequations}
where $\theta(x)$ is a unit-step function including Doppler shifts from drifting carriers,
$N_0(\Omega_{\rm ph})=[\exp(\hbar\Omega_{\rm ph}/k_BT)-1]^{-1}$ with $T$ as a lattice temperature and $\hbar\Omega_{\rm ph}$ as the longitudinal-optical-phonon energy,
while $\left|V^{\rm ep}_{k,k-q}\right|^2$ and $\left|V^{\rm hp}_{k',k'-q'}\right|^2$
are fully derived in Appendix\ \ref{app-3}.
If we neglect both ${\cal R}_{j,{\rm sp}}(k,\,t)$ terms in Eqs.\,\eqref{eq:dnedtREL} and \eqref{eq:dnhdtREL} and second terms in Eqs.\,\eqref{eq:Fe} and \eqref{eq:Fh}, as well as replace Boltzmann-type scattering terms in Eqs.\,\eqref{eq:dnedtSCAT} and \eqref{eq:dnhdtSCAT} by relaxation-time approximation,
we arrive at the linearized Boltzmann transport equations for electrons and holes.
\medskip

First, from the perspective of local quantum kinetics of carriers in 1D quantum wires, the DC-field induced photo-current density in each quantum wire is calculated as
\begin{equation}
\label{eq:dcj}
\mbox{\boldmath$J$}_{j,{\rm ph}}(t)
=\frac{e\alpha}{2\delta_0}\left[n_{j,{\rm 1D}}^{\rm h}(t)\,v^{\rm h}_j (t)-n_{j,{\rm 1D}}^{\rm e}(t)\,v^{\rm e}_j (t)\right]\mbox{\boldmath$\hat{e}$}_{\rm w}\equiv\left[\sigma_j^{\rm h}(t)+\sigma_j^{\rm e}(t)\right]
E_{\rm dc}\,\mbox{\boldmath$\hat{e}$}_{\rm w}\ ,
\end{equation}
where $\sigma_j^{\rm e(h)}(t)$ represent the quantum-wire transport conductivities of electrons and holes, the drift velocities $v_j^{\rm e(h)}(t)$ in Eq.\,\eqref{eq:dcj} are given by
\begin{equation}
\label{eq:vg}
v^{\rm e(h)}_j (t)=\frac{2}{N_j^{\rm e,h}(t)\hbar}\,\sum_{k}\frac{d\bar{E}^{\rm e(h)}_{j,k}(t)}{dk}\,n^{\rm e(h)}_{j,k}(t)\equiv\mu_j^{\rm e(h)}(t)\,E_{\rm dc}\ ,
\end{equation}
$\bar{E}^{\rm e(h)}_{j,k}(t)=\varepsilon_k^{\rm e(h)}+\Delta\varepsilon^{\rm e(h)}_{j,k}$ is the renormalized kinetic energy of electrons and holes,
and $\mu_j^{\rm e(h)}(t)$ are the quantum-wire nonlinear (with respect to $E_{\rm dc}$) mobilities of electron and holes.
Moreover, the local heating of electrons and holes in each quantum wire under a laser pulse can be described by their average kinetic energies per length:
\begin{equation}
\label{eq:thengy}
{\cal Q}^{\rm tot}_{j}(t)={\cal Q}^{\rm e}_{j}(t)+{\cal Q}^{\rm h}_{j}(t)\equiv \frac{2}{\cal L}\sum_k\,\bar{E}^{\rm e}_{j,k}(t)\,n_j^{\rm e}(t)+\frac{2}{\cal L}\sum_k\,\bar{E}^{\rm h}_{j,k}(t)\,n_j^{\rm h}(t)\ ,
\end{equation}
which can be used to determine the effective temperatures $T_{j,{\rm e(h)}}(t)$ for electrons and holes through the simple relations
${\cal Q}^{\rm e(h)}_{j}(t)=n_{j,{\rm 1D}}^{\rm e(h)}(t)k_{\rm B}T_{j,{\rm e(h)}}(t)/2$.
We can also calculate the time-resolved photoluminescence spectrum ${\cal P}_{j,{\rm pl}}(\Omega_0\,\vert\, t)$ for each quantum wire, given by\,\cite{huang-1,koch}
\begin{align}
\nonumber
\label{eq:pl}
{\cal P}_{j,{\rm pl}}(\Omega_0\,\vert\, t) &=
\frac{3d_{\rm cv}^2}{{\cal L}\epsilon_0  \sqrt{\epsilon_{\rm r}}}\,
\hbar \Omega_0 \,
\rho_0(\Omega_0)
\sum\limits_{k}\,n_{j,k}^{\rm e}(t)n_{j,k}^{\rm h}(t)\,
L(\hbar\Omega_0  	-  \varepsilon_{\rm G} - \varepsilon^{\rm e}_k
-  \varepsilon^{\rm h}_k  -  \varepsilon_{j,{\rm c}}(k,\,t), \; \hbar\gamma_{\rm eh})
\\
& \times
M(\hbar \Omega_0 - \varepsilon_{\rm G} - \varepsilon_{j,{\rm c}}(k=0,\,t), \; \hbar\gamma_{\rm eh})\ ,
\end{align}
where $\hbar\Omega_0$ is the energy of emitted photons.

\section{Electromagnetic Coupling in the Quantum Wires}
\label{interac}

From the solutions to Eq.\,\eqref{eq:dpdt}, we calculate the 1D polarization\,\cite{huang-3} introduced in Eq.\,\eqref{eq:Pq1}
\begin{subequations}
\begin{align}
\label{eq:qwpol}
%\color{red}
\tilde{\mbox{\boldmath$P$}}_{\rm qw}(\mbox{\boldmath$q$},t)
&
%\color{red}
=
\sum_{\sigma=x,y}\tilde{P}^\sigma_{\rm qw}(\mbox{\boldmath$q$},t)\,
\mbox{\boldmath$\hat{e}$}^\sigma_{\bf d}
=
\sum_j \texttt{e}^{-i {\bf q_\perp} \cdot {\bf R}^\perp_j  \,- \, q^2_\perp / 4\alpha^2 } \sum_{\sigma=x,y}
\tilde{{P}}^\sigma_{j}(q_\|,t)\,\mbox{\boldmath$\hat{e}$}^\sigma_{\bf d}\ ,
\\
\label{eq:Pq}
\tilde{{P}}^\sigma_{j}(q_\|,t)&=\frac{d_{\rm cv}\alpha}{2\delta_0{\cal L}}
\sum_{k} p^\sigma_{j, \, k+q_\|, \, k}(t) + {\rm H.C.}\ ,
\end{align}
\end{subequations}
where $\mbox{\boldmath$p$}_{j, \, k+q_\|, \, k}(t)$ is determined by Eq.\,\eqref{eq:dpdt},
$\hbar q_\|$ corresponds to the transferred momenta from photons to charged carriers in the wire direction,
${\cal L}$ is the quantum-wire length,
$\delta_0$ is the quantum-wire thickness,
$d_{\rm cv}$ is the 2D isotropic dipole moment between the valence and conduction band, and H.C. stands for the Hermitian conjugate term.
The free-charge density distribution in Eq.\,\eqref{eq:rhoq} is given by
\begin{equation}
\tilde{\rho}^{\rm 1D}_{j} (q_\|,t)
=\tilde{\rho}^{\rm h}_{j} (q_\|,t)  +  \tilde{\rho}^{\rm e}_{j} (q_\|,t)\ ,
\end{equation}
where $\tilde{\rho}^{\rm h}_{j} (q_\|,t)$ and  $\tilde{\rho}^{\rm e}_{j} (q_\|,t)$ are the charge-density distributions of holes and electrons in the $j$th quantum wire, which we calculate within the
random-phase approximation\,\cite{book-huang} as (see Appendix\ \ref{app1} for detailed derivations)
\begin{subequations}
\begin{align}
\label{eq:rhoh}
\tilde{\rho}_j^{\rm h}(q_\|,\,t)
& =
\frac{e\alpha}{N_j^{\rm e}(t){\cal L}\delta_0}
\sum_{k,k'}\sum_{\sigma=x,y}  \,  p^\sigma_{j, \, k', \, k-q_\|} (t)  \,  [p^\sigma_{j, \, k', \, k} (t)]^\ast
=
\frac{e\alpha}{N_j^{\rm e}(t){\cal L}\delta_0}
\sum_{k,k'}\, \mbox{\boldmath$p$}_{j, \, k', \, k-q_\|} (t)  \cdot [\mbox{\boldmath$p$}_{j, \, k', \, k} (t)]^\ast\ ,
\\
\label{eq:rhoe}
\tilde{\rho}_j^{\rm e}(q_\|,\,t)
& =
\frac{-e\alpha}{N_j^{\rm h}(t){\cal L}\delta_0}
\sum_{k,k'}\sum_{\sigma=x,y} \, [p^\sigma_{j, \, k - q_\|, \, k'} (t)]^\ast \, p^\sigma_{j, \, k , \, k'} (t)
=
\frac{-e\alpha}{N_j^{\rm h}(t){\cal L}\delta_0}
\sum_{k,k'}\,[\mbox{\boldmath$p$}_{j, \, k - q_\|, \, k'} (t)]^\ast \cdot \mbox{\boldmath$p$}_{j, \, k , \, k'} (t)\ .
\end{align}
\end{subequations}
Here, $N_j^{\rm e(h)}(t)=2\sum\limits_{k}\,n^{\rm e(h)}_{j,k}(t)$ is the total number of electrons (holes) in the $j$th quantum wire.
\medskip
	
Moreover, in Eqs.\,\eqref{eq:dnedt}-\eqref{eq:dpdt}, the renormalized Rabi frequencies can be calculated from
\begin{subequations}
\begin{align}
\nonumber
\Omega^x_{j,k,k^\prime}(t)
&=
\frac{d_{\rm cv}}{\color{black}\hbar}
\left.\int \frac{d\mbox{\boldmath$q$}_\perp}{\sqrt{q_\|^2+q_\perp^2}} \,
\left[-q_\|\tilde{\cal E}_{j,x}^{\perp}(\mbox{\boldmath$q$}_\perp,q_\|,\,t)+q_\perp\tilde{\cal E}_{j,x}^{\|}(\mbox{\boldmath$q$}_\perp,q_\|,\,t)\right]\right|_{q_\|=k-k^\prime}
\\
\label{eq:Rabix}
&+ \sum_{k_1\ne k,\,k'_1\ne k'}  \,   p^x_{j,k_1,k'_1}(t)   \,   V^{\rm eh}_{k,k^\prime;\,k_1^\prime,k_1}\ ,
\\
\nonumber
\Omega^y_{j,k,k^\prime}(t)
&=
\frac{d_{\rm cv}}{\color{black}\hbar}
\left.\int \frac{d\mbox{\boldmath$q$}_\perp}{\sqrt{q_\|^2+q_\perp^2}} \,
\left[q_\perp\tilde{\cal E}_{j,y}^{\perp}(\mbox{\boldmath$q$}_\perp,q_\|,\,t)+q_\|\tilde{\cal E}_{j,y}^{\|}(\mbox{\boldmath$q$}_\perp,q_\|,\,t)\right]\right|_{q_\|=k-k^\prime}
\\
\label{eq:Rabiy}
&+ \sum_{k_1\ne k,\,k'_1\ne k'}  \,   p^y_{j,k_1,k'_1}(t)   \,   V^{\rm eh}_{k,k^\prime;\,k_1^\prime,k_1}\ ,
\end{align}
\end{subequations}
where $\mbox{\boldmath$q$}=\{q_\perp\hat{\mbox{\boldmath$e$}}_x,\,q_\|\hat{\mbox{\boldmath$e$}}_y\}$ for the chosen
$\hat{\mbox{\boldmath$e$}}_{\rm w}=\hat{\mbox{\boldmath$e$}}_y$,
$\hat{\mbox{\boldmath$e$}}_{\perp{\bf q}}=\hat{\mbox{\boldmath$e$}}_z\times\hat{\mbox{\boldmath$e$}}_{\bf q}$ ($\hat{\mbox{\boldmath$e$}}_z$ is a unit
vector in the direction perpendicular to the $xy$-plane), and the second
term represents the correction to the dipole moment by excitonic interactions.
The effective transverse and longitudinal electric-field components, $\tilde{\cal E}_{j,x(y)}^{\perp}(\mbox{\boldmath$q$}_\perp,\mbox{\boldmath$q$}_\|,\,t)$ and $\tilde{\cal E}_{j,x(y)}^{\|}(\mbox{\boldmath$q$}_\perp,\mbox{\boldmath$q$}_\|,\,t)$,
are the 1D finite Fourier-transformed
corresponding electric-field vectors inside the $j$th wire:
\begin{subequations}
\begin{align}
\label{eq:tfield}
\mbox{\boldmath$\tilde{{\cal E}}$}_j^{\perp}(\mbox{\boldmath$q$}_\perp,q_\|,\,t)
&=
\int\limits_{-\infty}^{\infty} \frac{dr_\|}{\cal L} \,\texttt{e}^{-iq_\|r_\|} \,g(r_\|)  \int\limits_{-\infty}^{\infty} d\boldsymbol{r}_\perp\,
\texttt{e}^{-i{\bf q}_\perp\cdot \boldsymbol{r}_\perp}
\psi^{\rm e}_0(\boldsymbol{r}_\perp-\boldsymbol{R}^\perp_j)\psi^{\rm h}_0(\boldsymbol{r}_\perp-\boldsymbol{R}^\perp_j)\,
\mbox{\boldmath${E}$}^\perp(\boldsymbol{r}_\perp, r_\|,\, t)\ ,
\\
\label{eq:lfield}
\mbox{\boldmath$\tilde{{\cal E}}$}_j^{\|}(\boldsymbol{q}_\perp,q_\|,\,t)
&=
\int\limits_{-\infty}^{\infty} \frac{dr_\|}{\cal L} \,\texttt{e}^{-iq_\|r_\|} \, g(r_\|)\int\limits_{-\infty}^{\infty} d\boldsymbol{r}_\perp \,
\texttt{e}^{-i{\bf q}_\perp \cdot \boldsymbol{r}_\perp}
\psi^{\rm e}_0(\boldsymbol{r}_\perp-\boldsymbol{R}^\perp_j)\psi^{\rm h}_0(\boldsymbol{r}_\perp-\boldsymbol{R}^\perp_j)\,
\mbox{\boldmath${E}$}^\|(\boldsymbol{r}_\perp, r_\|,\, t)\ .
\end{align}
\end{subequations}
Here, we take $g(x) = \Gamma(9/8) \exp \left[  - (2 x / {\cal L})^8   \right]$ as a normalized gating function for the wire of length ${\cal L}$.
\medskip

In addition, from the constraint $\partial\tilde{\mbox{\boldmath$D$}}^\|(\mbox{\boldmath$q$},t)/\partial t+\tilde{\mbox{\boldmath$J$}}^\|_{\rm qw}(\mbox{\boldmath$q$},t)=0$ and Eq.\,\eqref{eq:ME1}, the current
$\tilde{J}^{\rm 1D}_j(q_\|,t)$ in the $j$th
quantum wire is found to be
\begin{equation}
\label{eq:qwj}
\tilde{J}^{\rm 1D}_j(q_\|,t)=\frac{\tilde{J}^{\|}_{j}(\mbox{\boldmath$q$},t)}{(\mbox{\boldmath$\hat{e}$}_{\rm w}\cdot\mbox{\boldmath$\hat{e}$}_{\bf q})}
=-\frac{1}{iq(\mbox{\boldmath$\hat{e}$}_{\rm w}\cdot\mbox{\boldmath$\hat{e}$}_{\bf q})}\,\frac{\partial}{\partial t}\left[ \tilde{\rho}_j^{\rm h}(q_\|,\,t)+\tilde{\rho}_j^{\rm e}(q_\|,\,t)\right]
=\frac{i}{q_\|}\,\frac{\partial}{\partial t}\left[ \tilde{\rho}_j^{\rm h}(q_\|,\,t)+\tilde{\rho}_j^{\rm e}(q_\|,\,t)\right]\ ,
\end{equation}
where we treat a quantum wire as a quasi-1D electronic system in the electric quantum limit with a current flowing only along the $\mbox{\boldmath$\hat{e}$}_{\rm w}$ direction.
From Eqs.\,\eqref{eq:Pq}, \eqref{eq:rhoh} and \eqref{eq:rhoe}, we know that Eq.\,\eqref{eq:qwj} has provided us with a constant in time from the dynamical equation with respect to $p^\sigma_{j,k,k^\prime}(t)$,
which can be employed to determine both the transient and steady-state optical response of individual photo-excited quantum wire. For steady state, however, we can simply replace the occupations
$n^{\rm e}_{j, k}(t)$ and $n^{\rm h}_{j, k'}(t)$ by their thermal-equilibrium Fermi functions $1/\{1+\exp[(\varepsilon^{\rm e}_{k}-\mu_{\rm e})/k_BT]\}$ and $1/\{1+\exp[(\varepsilon^{\rm h}_{k'}-\mu_{\rm h})/k_BT]\}$,
respectively, where $\mu_{\rm e}$ and $\mu_{\rm h}$ are the chemical potentials of electrons and holes, and $T$ is the lattice temperature.
Meanwhile, it implies a conservation law, i.e., the charge conservation law.
Moreover, the left-hand side term, $\tilde{J}^{\rm 1D}_j(q_\|,t)$, can be computed perturbatively for weak fields by using a linear-response theory\,\cite{book-huang} (i.e., the Kubo formula) to obtain conductivities,
while the right-hand side term, $\partial^2\tilde{\rho}_j^{\rm e,h}(q_\|,\,t)/\partial t^2$, can be treated by using the random-phase approximation\,\cite{book-huang} for high carrier densities to find plasmon frequencies.
\medskip

Finally, from the perspective of propagation of incident pulsed light ${\mbox{\boldmath$E$}}^\perp_{{\rm inc}}(\mbox{\boldmath$r$},t\,\vert\, \omega_{0})$,
we can compute its coherent Fourier spectra for intensity transmission $\mathbbmsl{T}_{{\rm F}}(\Omega\,\vert\,\omega_{0})$
and reflection $\mathbbmsl{R}_{{\rm F}}(\Omega\,\vert\,\omega_{0})$, i.e., transient wavefront detection at a fixed spatial position, as functions of Fourier frequency $\Omega$, given by
\begin{subequations}
\begin{align}
\label{trans}
\mathbbmsl{T}_{{\rm F}}(\Omega\,\vert\,\omega_{0})&=
\frac{\int\limits_{-\infty}^{+\infty} d\mbox{\boldmath$r$}_\perp\left|\mbox{\boldmath$E$}^\perp(\mbox{\boldmath$r$}_\perp,r_\|\gg{\cal L}/2,\Omega\,\vert\,\omega_{0})\right|^2}
{{\cal W}\left({\cal E}_{0}^{\rm inc}\right)^2}\ ,
\\
\label{refle}
\mathbbmsl{R}_{{\rm F}}(\Omega\,\vert\,\omega_{0})&=
\frac{\int\limits_{-\infty}^{+\infty} d\mbox{\boldmath$r$}_\perp\left|\mbox{\boldmath$E$}^\perp(\mbox{\boldmath$r$}_\perp,r_\|\ll-{\cal L}/2,\Omega\,\vert\,\omega_{0})\right|^2}
{{\cal W}\left({\cal E}_{0}^{\rm inc}\right)^2}\ ,
\end{align}
\end{subequations}
where ${\mbox{\boldmath$E$}}^\perp(\mbox{\boldmath$r$},\Omega\,\vert\,\omega_{0})$ is the Fourier transform of
$\mbox{\boldmath$E$}^\perp(\mbox{\boldmath$r$},t\,\vert\,\omega_{0})$ with respect to $t$, $\omega_0$ is the central frequency of the incident light pulse, and
$r_{\|}^{(0)}\ll-{\cal L}/2$ is the peak position of initial incident light pulse at $t=0$.
Moreover, ${\cal E}_{0}^{\rm inc}$ is the amplitude of the incident light pulse and ${\cal W}$ represents the width of the quantum-wire array.

\section{Simulation Results and Discussions}
\label{sandm}

\begin{figure}%[p]
\centering
\includegraphics[width=0.4\textwidth]{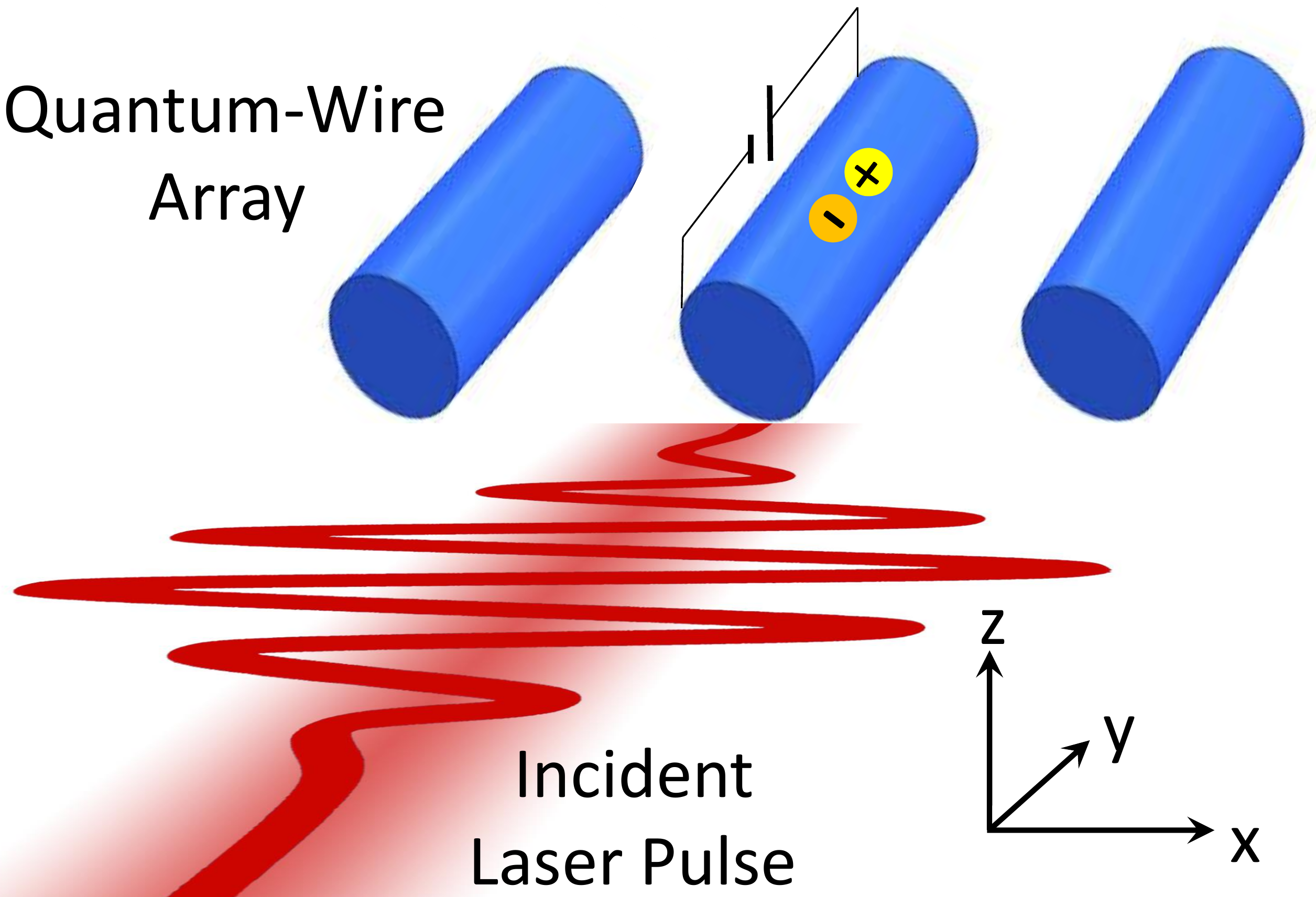}
\caption{(Color online) Schematic of a model system which consists of biased quantum wires extending along the $y$ direction and displayed in the $x$ direction by a linear array.
An incident laser pulse, with a Gaussian spatial profile in the $x$ direction and its electric and magnetic fields along $x$ and $z$ directions, propagates along the $y$ direction and generates e-h pairs in quantum wires by interacting with them.
Additionally, induced electrons and holes in quantum wires are driven by DC electric fields.}
\label{illustration}
\end{figure}

We numerically solve Maxwell's equations in both 1D and 2D systems for a
$\tau_0 = 40\,$fs (full width at half maximum), $\lambda_0=800\,$nm wavelength ($\omega_0=2\pi c/\lambda_0\equiv q_0c$)
pulsed-laser field propagating in the $y$-direction. The magnetic field is polarized purely in the $z$-direction.
The corresponding incident electric field is primarily polarized in the $x$-direction, but also has a significant $y$-component in the 2D spatial simulations because of tight initial focusing (initial beam width is taken to be $w_x=800\,$nm for the 2D case)
and the light diffraction by a linear array of quantum wires embedded in a dielectric host.
The initial peak intensity of the pulse is $6.2\,$GW/cm$^2$.
The laser field immediately propagates through an AlAs host material.  The linear polarization in the AlAs host is calculated by a Lorentz model, for which the necessary constants are calculated from a Sellmeier equation
for AlAs\,\cite{doi:10.1063/1.1660760}. The AlAs index of refraction at the peak wavelength is given by $n_0=3.0044$. The pulse propagates toward the quantum-wire array that is centered about $x=0$ and $y=0$, as illustrated in Fig.\,\ref{illustration}.
\medskip

The initial magnetic field is numerically constructed as a diffracting (dispersive) Gaussian beam (pulse)\cite{diels}:
\begin{equation}
\label{pulse}
\mbox{\boldmath$H$}(x,y,t=0) = \hat{\mbox{\boldmath$e$}}_z\,H_{z0}\, e^{i k_0 (y-y_0)}
                  \exp\left\{-\frac{[1+ i b(y-y_0)]x^2}{\ell_x^2(y-y_0,L_R)}\right\}\,
                  \exp\left\{\frac{-[1- i  a(y-y_0)](y-y_0)^2}{\ell_y^2(y-y_0, L_D)}\right\}\ ,
\end{equation}
where $y_0$ is the initial $y$-position of the pulse peak and $w_y = (c/n_0)\tau_0/\sqrt{2\ln2}$ is the initial pulse length.
The wave vector at the peak wavelength is $k_0=2\pi n_0/\lambda_0$ while the initial peak magnetic field $H_{z0}$.  The functions $a(y)=y/L_D$, $b(y)=y/L_R$, and $\ell_{x,y}(y,L) = w_{x,y}\sqrt{1+(y/L)^2}$,
where $L_D= k_0 w_y^2 / 2$ is the host dispersion length and $L_R = k_0 w_x^2 / 2$ is the Rayleigh range. For the 1D case, we have $x=0$.
\medskip

%%%%%%%%%%%%%%%%%%%%%%%%%%%%%%%%%%%%%%%%%%%%%%%%
\begin{table}[htbp]
	\centering
	\caption{\bf Parameters for AlAs host semiconductor}
	\begin{tabular}{cccc}
		\hline
		Parmeter 					&	Description										& 	Value	& 	Units 			\\
		\hline
		$\epsilon_s$				&	Static dielectric constant					& 10.0		&						\\
		$\epsilon_\infty$		&	High-frequency  constant					& 8.2			&						\\
		$\epsilon_r$ 				&	Relative dielectric constant 				& 9.1			&						\\
		$\Omega_{\rm ph}$	&	Phonon frequency								&	36			& meV/$\hbar$	\\
		$\Gamma_{\rm ph}$	&	Inverse phonon lifetime						&	1			& meV/$\hbar$	\\
		$T$							&	Host temperature								&	77			& K					\\
		\hline
	\end{tabular}
	\label{tab:host}
\end{table}
\begin{table}[htbp]
	\centering
	\caption{\bf Parameters for GaAs quantum wires}
	\begin{tabular}{cccc}
		\hline
		Parmeter 				&	Description	& 	Value	& 	Units \\
		\hline
		$\cal{L}$					&	Length of quantum wire			&	200					&	nm				\\
		$\delta_0$				&	Thickness of quantum wire		&	5.65					&	nm				\\
		$\hbar \Omega_0$	&	Energy level separation 		&	100					&	meV 				\\
		$\varepsilon_{\rm G}$	&	Band gap								&	1.5					&	eV				\\
		$m^*_e  $					&	Electron effective mass 		&	0.07\,$m_0$	&  kg					\\
		$m^*_h  $					&	Hole effective mass 				&	0.45\,$m_0$	&	kg					\\
		$\gamma_e$				&	Electron lifetime frequency	&	20						&	THz				\\
		$\gamma_h$				&	Hole lifetime frequency			&	20						&	THz				\\
		$E_{\rm dc}$				&	Applied DC field					&	$1$				& kV/cm				\\
		\hline
	\end{tabular}
	\label{tab:wire}
\end{table}
%%%%%%%%%%%%%%%%%%%%%%%%%%%%%%%%%%%%%%%%%%%%%%%%

To calculate the corresponding initial electric-field vector, we first choose the gauge $\mbox{\boldmath$\nabla$}\cdot \mbox{\boldmath$A$} = 0$, allowing us to first calculate the magnetic vector potential $\mbox{\boldmath$A$}$ and then the electric field by
\begin{equation}
\tilde{\mbox{\boldmath$A$}} ({\mbox{\boldmath$q$}},t=0) =
-i\left(\frac{\mbox{\boldmath$q$}}{\mu_0q^2} \right)
\times
\tilde{\mbox{\boldmath$H$}} ({\mbox{\boldmath$q$}},t=0)\ ,
\nonumber
\end{equation}
\begin{equation}
\tilde{\mbox{\boldmath$E$}} ({\mbox{\boldmath$q$}},t=0) =
- \left[
\frac{\partial \tilde{\mbox{\boldmath$A$}}({\mbox{\boldmath$q$}},t)}{\partial t}
\right]_{t=0}
=
- \left(\frac{\mbox{\boldmath$q$}\,\omega_q}{\mu_0q^2}\right)
\times
\tilde{\mbox{\boldmath$H$}} ({\mbox{\boldmath$q$}},t=0)\ .
\end{equation}
Here, the initial time dependance of the fields is assumed to be given by $\exp{(- i \omega_q t)}$, where $\omega_q$ is the $q$-dependent frequency obtained by solving $|\mbox{\boldmath$q$}| = n_r(\omega_q)\,\omega_q / c$ and $n_r(\omega)$ is the
frequency-dependent refractive index in the host.  The magnetic field is first propagated one-half time step by multiplying $\tilde{\mbox{\boldmath$H$}}(\mbox{\boldmath$q$},t=0)$ with $\exp{(- i \omega_q \Delta t/2)}$, then all fields are inverse Fourier transformed back into $xy$-space where their real part is taken.  After Fourier transforming back into $\mbox{\boldmath$q$}$-space, we use the standard PSTD method to propagate the pulse into the quantum-wire array.
The properties of the host material used in the quantum-wire calculations in Sec.\,\ref{sec:SBE} are summarized in Table\,\ref{tab:host}.
The quantum wires themselves are assumed to be made of GaAs and their properties are summarized in Table\,\ref{tab:wire}. Each wire is oriented along the $y$ direction ($\mbox{\boldmath$\hat{e}$}_{\rm w}=\mbox{\boldmath$\hat{e}$}_{y}$),
such that $\mbox{\boldmath$q$}_\| = q_y\mbox{\boldmath$\hat{e}$}_y$, $\mbox{\boldmath$q$}_\perp = q_x\mbox{\boldmath$\hat{e}$}_x$, $\xi_\| = y$, and $\mbox{\boldmath$\xi$}_\perp = x \mbox{\boldmath$\hat{e}$}_x$.  The linear array of quantum wires is centered about $x=0$ and $y=0$, and each wire is separated by a distance of $a$ in a linear array. The 2D simulations are performed with arrays of 1, 3, and 10 wires.

\begin{figure*}%[p]
\centering
\includegraphics[width=0.8\textwidth]{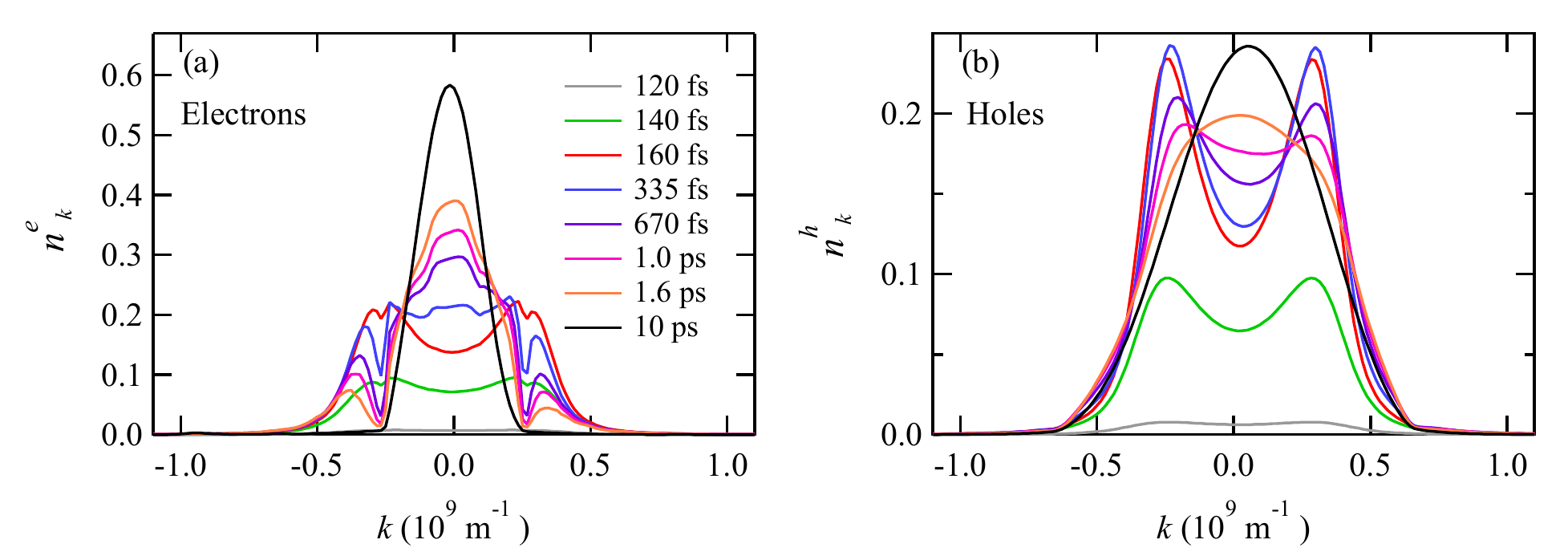}
\caption{(Color online) Calculated $n^{\rm e}_{j,k}(t)$ [in ($a$)] and $n^{\rm h}_{j,k}(t)$ [in ($b$)] from Eqs.\,(\ref{eq:dnedt}) and (\ref{eq:dnhdt})
as functions of carrier wave number $k$ for electrons (e) and holes (h) within the central quantum wire at different times $t$. Here, results for electrons and holes are shown for different moments at
$t=120,\,140,\,160,\,335,\,670\,$fs and $t=1,\,1.6,\,10\,$ps for a $40\,$fs light pulse interacting significantly with electrons in quantum wires within the time interval $t\in100-180\,$fs.}
\label{occupation}
\end{figure*}

\subsection{Transient Quantum Electronic Properties}
\label{disc-1}

The experimentally-measurable field and optical responses of solid-state materials can be computed quantum-statistically using non-equilibrium occupations for different electronic states.
In Fig.\,\ref{occupation}, by solving Eqs.\,\eqref{eq:dnedt} and \eqref{eq:dnhdt} we present comparisons of $n^{\rm e}_{j,k}(t)$ [in (a)] and $n^{\rm h}_{j,k}(t)$ [in (b)]
as functions of carrier wave number $k$ for electrons (e) and holes (h) within the central quantum wire at different times ($t$).
Here, the occupations $n^{\rm e,h}_{j,k}(t)$ are slightly asymmetric with respect to $k=0$ due to the presence of a $1\,$kV/cm DC electric field $E_{\rm dc}$, and
the time-evolutions of non-equilibrium hot electron and hole distributions are displayed with a resonant emission of longitudinal-optical phonons by electrons (two dips on tails).
As the pulse just reaches the quantum wire ($t=120\,$fs),
very weak stimulated absorption occurs first. Electrons are promoted from the lower valence subband to the upper conduction subband, leaving holes behind in the valence subband. Such a coherent process
appears as double peaks in both $n^{\rm e}_{j,k}(t)$ and $n^{\rm h}_{j,k}(t)$, which are almost identical as the pulse maximum sits inside the quantum wire ($t=140,\,160\,$fs).
After the pulse leaves the
quantum wire ($t=335\,$fs), significant effects from electron-electron and hole-hole scattering show up. As a result, the double-peak
occupations are replaced by one sharp (electron) and one round (hole) peaks ($t=670\,$fs).
These two inequivalent non-thermal processes lead to much hotter electrons than holes ($t=1,\,1.6,\,10\,$ps). Meanwhile, a {\em phonon-hole burning},
which is completely different from the well-known spectral-hole burning\,\cite{hole-burning} at the pumping resonance $\varepsilon^{\rm e}_k+\varepsilon^{\rm h}_k=\hbar\omega_0-\varepsilon_{\rm G}$ in a gain spectrum
because of dominant stimulated emission,
develops in $n^{\rm e}_{j,k}(t)$ but not in $n^{\rm h}_{j,k}(t)$ due to resonant emission of longitudinal-optical phonons by electrons ($t\in 335\,$fs$-1.6\,$ps).
This phonon-hole burning process is accompanied by a rising central peak in $n^{\rm e}_{j,k}(t)$ due to energy relaxation of hot electrons to the $\varepsilon^{\rm e}_{k}=0$ subband edge with a very large density-of-states for the quantum wire.
As time further goes well beyond $t\gg 1.6\,$ps, this phonon-hole burning will gradually disappear as more and more high-energy electrons relax to lower energies, ending with a high
round peak surrounded by two smooth tails on each side (i.e., quasi-thermal-equilibrium distribution) but still giving rise to a much higher electron temperature than that of holes.
From Fig.\,\ref{occupation}(a), we conclude that an initial quasi-thermal-equilibrium state starts forming for electrons at $t=670\,$fs
after the light pulse has passed through the quantum wires, which gives rise to an initial thermalization time $t_0\sim 670\,$fs.

\begin{figure*}%[p]
\centering
\includegraphics[width=0.8\textwidth]{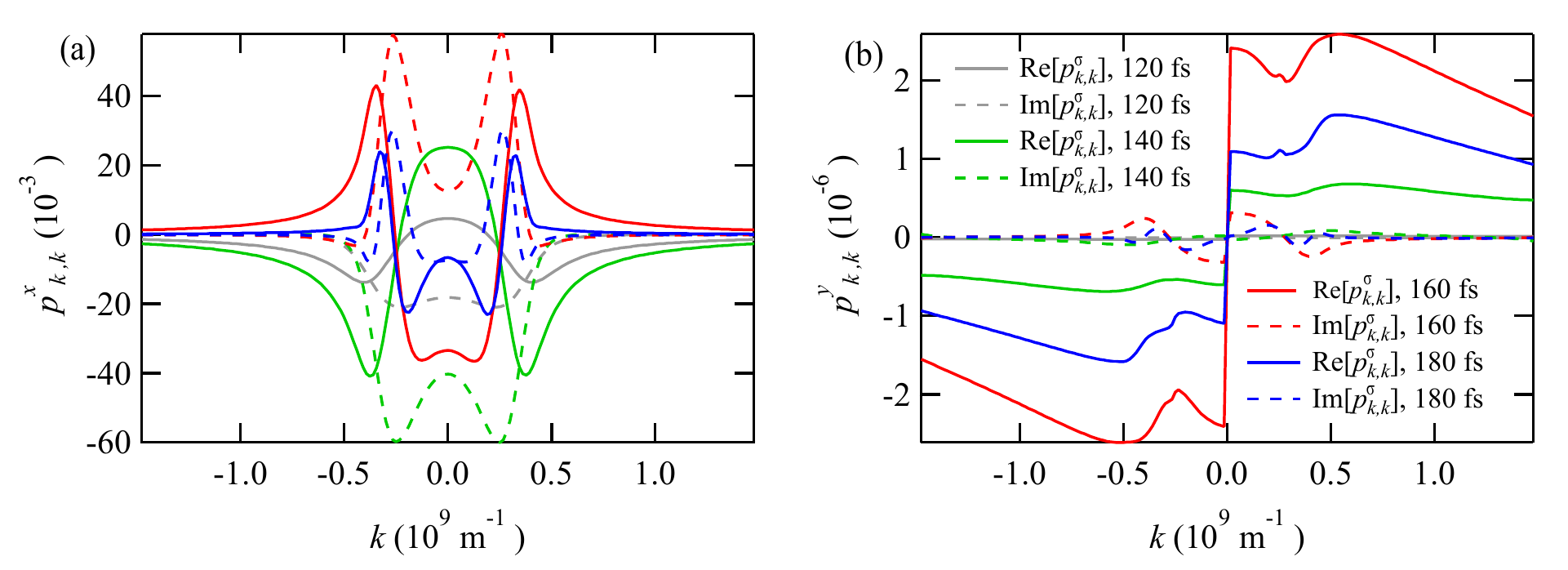}
\caption{(Color online) Calculated ${\rm Re}[p^\sigma_{j,k,k'}(t)]$ (solid) and ${\rm Im}[p^\sigma_{j,k,k'}(t)]$ (dashed) from Eq.\,(\ref{eq:dpdt}) as functions of carrier wave number
$k$ for optical coherence of electron-hole pairs within the central quantum wire at different moments at
$t=120,\,140,\,160,\,180\,$fs for a $40\,$fs light pulse interacting significantly with electrons in quantum wires within the time interval $t\in100-180\,$fs.  Here, results for $\sigma=x,y$,
corresponding to two directions perpendicular and parallel to quantum wires, are shown
in ($a$) with $k'=k$ and ($b$) with $k'=q_0$, respectively.}
\label{coherence}
\end{figure*}

\medskip

Physically speaking, the change in non-equilibrium occupations is attributed to an optical response (or induced optical coherence) of photo-excited carriers, which leads to light coupling between upper conduction and lower valence
subbands of electrons in semiconductor materials.
In Fig.\,\ref{coherence}(a), by solving Eq.\,\eqref{eq:dpdt}
we present comparisons of both ${\rm Re}[p^x_{j,k,k}(t)]$ (solid) and ${\rm Im}[p^x_{j,k,k}(t)]$ (dashed) for transverse optical coherences of induced electron-hole pairs
as functions of carrier wave number $k$ within the central quantum wire at different times.
Here, the time-evolution of negative double peaks in ${\rm Im}[p^x_{j,k,k}(t)]$ ($t=120\,$fs) for initial resonant stimulated emission at finite $|k|$ values is demonstrated in the presence of incident light pulse
with $\hbar\omega_0>\varepsilon_{\rm G}$.
For vertical electron transitions with $k=k'$, we find in a perturbative way that
$p^x_{j,k,k}(t)\sim -\texttt{e}^{-(i/\hbar)(\bar{E}^{\rm e}_{j,k}+\bar{E}^{\rm h}_{j,k}+\varepsilon_{\rm G})t}/[\hbar(\omega_0+i\gamma_{\rm eh})-(\bar{E}^{\rm e}_{j,k}+\bar{E}^{\rm h}_{j,k}+\varepsilon_{\rm G})]$.
Therefore, both ${\rm Re}[p^x_{j,k,k}(t)]$ and ${\rm Im}[p^x_{j,k,k}(t)]$ become even functions of $k$, and
we expect a sign switching in both ${\rm Re}[p^x_{j,k,k}(t)]$ and ${\rm Im}[p^x_{j,k,k}(t)]$ themselves (comparing results at $t=140,\,160\,$fs)
as $(\bar{E}^{\rm e}_{j,k}+\bar{E}^{\rm h}_{j,k}+\varepsilon_{\rm G})t/\hbar$
varies from $2\ell\pi$ to $(2\ell+1)\pi$ (for {\rm Re}) or from $(2\ell+1/2)\pi$ to $(2\ell+3/2)\pi$ (for {\rm Im}), where $\ell$ is an integer.
Here, positive (negative) double peaks in ${\rm Im}[p^x_{j,k,k}(t)]$ imply a stimulated absorption (emission), i.e.,
coherent interband Rabi oscillations of electrons.
Moreover, there exist vanishing stimulated transitions at two specific $k$ values due to ${\rm Re}[p^x_{j,k,k}(t)]=0$
at $\bar{E}^{\rm e}_{j,k}+\bar{E}^{\rm h}_{j,k}=\hbar\omega_0-\varepsilon_{\rm G}$. Meanwhile, the zero transition at this moment
is further accompanied by weak stimulated emission for large $|k|$ values and strong stimulated absorptions for small $|k|$ values at $t=160\,$fs.
\medskip

In addition to transverse optical coherence $p^x_{j,k,k}(t)$ which can affect light propagation far away from quantum wires, the laser pulse also introduces a longitudinal optical coherence $p^y_{j,k,k'}(t)$
due to induced longitudinal-plasma waves oscillating along the wire direction.
In Fig.\,\ref{coherence}(b), we display comparisons of both ${\rm Re}[p^y_{j,k,q_0}(t)]$ (solid) and ${\rm Im}[p^y_{j,k,q_0}(t)]$ (dashed) for longitudinal responses from these induced plasma waves
as functions of carrier wave number $k$ within the central quantum wire at different $t$.
For non-vertical electron transitions with $k\neq k'$, we find in a similar way that
$p^y_{j,k,k'}(t)\sim{\rm sgn}(k-k')\,\hbar\Omega^y_{j,k,k'}(t)[1-n^{\rm e}_{j,k}(t)-n^{\rm h}_{j,k'}(t)]/[\hbar(\omega_0+i\gamma_{\rm eh})-(\bar{E}^{\rm e}_{j,k}+\bar{E}^{\rm h}_{j,k'}+\varepsilon_{\rm G})]$ with
${\rm sgn}(x)$ as a sign function.
Therefore, both ${\rm Re}[p^y_{j,k,q_0}(t)]$ and ${\rm Im}[p^y_{j,k,q_0}(t)]$ appear approximately as odd functions of $k$ with a sharp sign switching for ${\rm Re}[p^y_{j,k,q_0}(t)]$ at $k=q_0\approx 0$.
Here, large ${\rm Re}[p^y_{j,k,q_0}(t)]$ represents a significant nonlocal ($k$-dependent) correction to the dielectric constant of quantum wires from contributions of photo-excited free carriers,
while small ${\rm Im}[p^y_{j,k,q_0}(t)]$ indicates a weak light-induced optical current flowing within the quantum wire.
Furthermore, nonzero ${\rm Im}[p^y_{j,k,q_0}(t)]$ implies a finite lifetime for induced plasma waves and oscillations of ${\rm Im}[p^y_{j,k,q_0}(t)]$ with $k$ correspond to
dissipation (positive values) and amplification (negative values) of plasma waves due to their energy exchange with the laser pulse.

\begin{figure}%[p]
\centering
\includegraphics[width=0.45\textwidth]{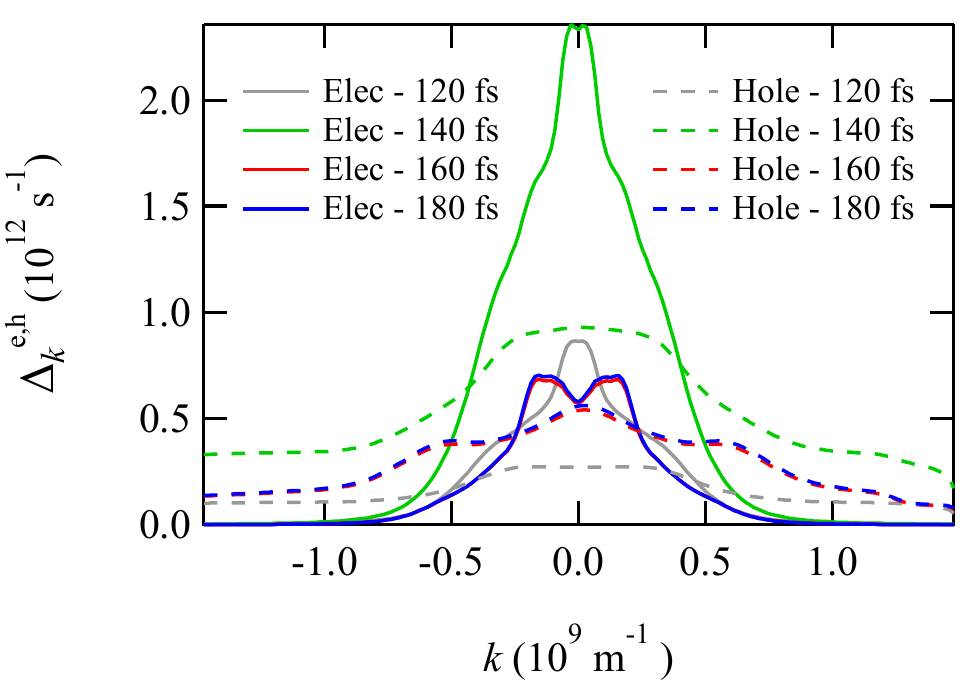}
\caption{(Color online) Calculated diagonal dephasing rates for electrons $\Delta^{\rm e}_{j,k}(t)$ (solid) and holes $\Delta^{\rm h}_{j,k}(t)$ (dashed)
from Eqs.\,\eqref{diage} and \eqref{diagh} are presented as functions of wave number $k$ for electrons and holes within the central quantum wire,
where results for electrons and holes are shown for different moments at $t=120,\,140,\,160,\,180\,$fs for a $40\,$fs light pulse interacting with quantum wires within the time interval $t\in 100-180\,$fs.}
\label{dephasingd}
\end{figure}

\medskip

The induced optical coherence $\mbox{\boldmath$p$}_{j,k,k'}(t)$ of photo-excited carriers in quantum wires suffers from a decay with time (i.e., optical dephasing) due to carrier scattering with phonons and other carriers,
and the dephasing rate characterizes how fast an excited-state configuration (or photon quantum memory) by incident laser pulse will elapse with time.
In Fig.\,\ref{dephasingd}, we compare diagonal-dephasing rates of induced quantum coherence in Fig.\,\ref{coherence}, for
both electrons $\Delta^{\rm e}_{j,k}(t)$ (solid) and holes $\Delta^{\rm h}_{j,k}(t)$ (dashed) as functions of carrier wave number $k$ within the central quantum wire at different times.
Here, $\Delta^{\rm e}_{j,k}(t)$ and $\Delta^{\rm h}_{j,k}(t)$ are being built up as the light pulse enters into the quantum wire ($t=120,\,140\,$fs), with a sharp and a round peak at $k=0$
for electrons and holes, respectively, due to very small electron occupation $n^{\rm e}_{j,k}(t)$ at $k=0$ for the final state in pair-scattering processes.
After the pulse maximum moves into the quantum wire ($t=160,\,180\,$fs),
a single peak in $\Delta^{\rm e}_{j,k}(t)$ has been replaced by double peaks. However, the dip between two peaks is absent in $\Delta^{\rm h}_{j,k}(t)$
due to a relatively large broadening effect on pair scattering between heavier holes.
The dual-peak structure associated with $\Delta^{\rm e}_{j,k}(t)$ is reminiscent of the corresponding feature in $n^{\rm e}_{j,k}(t)$, as shown in Fig.\,\ref{occupation}(a).
Furthermore, the dual-peak structure in $\Delta^{\rm e}_{j,k}(t)$ is accompanied by significant reductions of peak strengths of $\Delta^{\rm e,h}_{j,k}(t)$, which are attributed to enhanced Pauli-blocking effects on final states in
pair scattering of electrons and holes as occupations $n^{\rm e,h}_{j,k}(t)$ at $k=0$ are greatly increased.
Such a Pauli-blocking effect is enhanced greatly due to resonant emission of longitudinal-optical phonons for electron transitions down to $k=0$ state.

\begin{figure}%[p]
\centering
\includegraphics[width=0.5\textwidth]{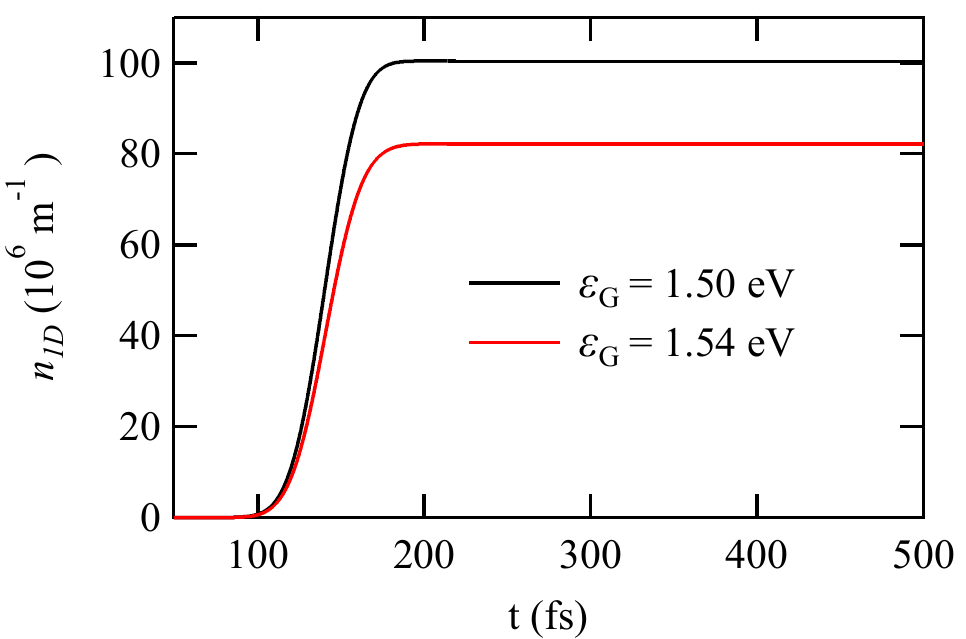}
\caption{(Color online) Calculated photo-generated carrier density $n_{\rm j,1D}(t)=n_{j,{\rm 1D}}^{\rm e}(t)=n_{j,{\rm 1D}}^{\rm h}(t)=N_j^{\rm e,h}(t)/{\cal L}$ [see its expression right after Eqs.\,(\ref{eq:rhoh}) and (\ref{eq:rhoe})]
as a function of $t$ for electrons (e) and holes (h) within the central quantum wire for the bandgap $\varepsilon_{\rm G}=1.50\,$eV (black) and $\varepsilon_{\rm G}=1.54\,$eV (red).
Here, a $40\,$fs light pulse interacts significantly with quantum wires within the time interval $t\in100-180\,$fs.}
\label{density}
\end{figure}
\begin{figure}%[p]
\centering
\includegraphics[width=0.5\textwidth]{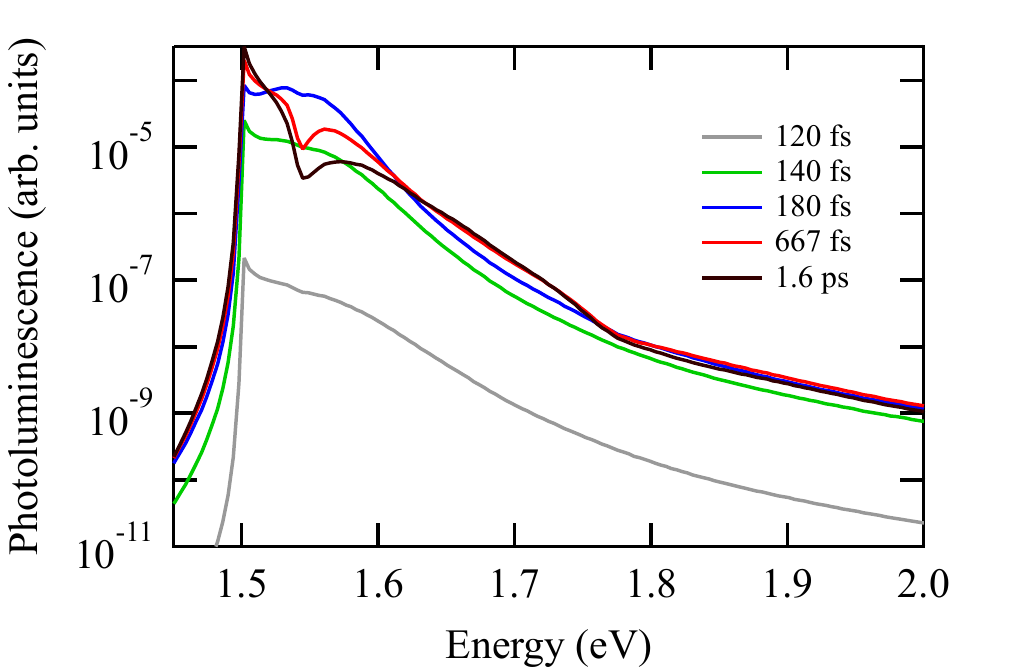}
\caption{(Color online) Calculated time-evolution of photoluminescence spectra ${\cal P}_{j,{\rm pl}}(\Omega_0\,\vert\, t)$ (logarithmic scale) from Eq.\,(\ref{eq:pl}) for spontaneous emission within the central quantum wire at different times
$t=120,\,140,\,180,\,667\,$fs and $t=1.6\,$ps. Here, $\hbar\Omega_0$ is the energy of spontaneously emitted photons and a $40\,$fs light pulse interacts significantly with quantum wires within the time interval $t\in100-180\,$fs.}
\label{luminescence}
\end{figure}

\medskip

In a quantum-statistical theory, different electrons in a system can be labeled by their individual electronic states or wave number $k$ (including spin degeneracy).
The total number of electrons can be found by summing all $k$-dependent occupations with respect to $k$.
In Fig.\,\ref{density}, after using obtained occupations $n^{\rm e,h}_{j,k}(t)$
we present the photo-generated carrier density $n_{j,{\rm 1D}}^{\rm e}(t)=n_{j,{\rm 1D}}^{\rm h}(t)\equiv n_{j,{\rm 1D}}(t)$ as functions of time $t$ for electrons and holes within the central quantum wire. As the pulse maximum reaches the quantum wire
($t<140\,$fs), the carrier density increases quickly with $t$. Soon after the pulse passes the quantum wire ($t\approx200\,$fs),
the carrier density reaches a peak value and becomes nearly constant for $t>200\,$fs.
This feature results from the fact that the spontaneous emission ${\cal R}_{j,{\rm sp}}(k,\,t)$ is insignificant on this time scale, and therefore the total number of photo-generated carriers is conserved after the pulse tail has left the quantum wire.
However, on this time scale the electron and hole non-thermal occupations, as functions of $k$, still change dramatically with $t$ due to very strong Coulomb and optical-phonon scattering of carriers within their individual subbands.
Such carrier scattering processes eventually lead to achieving quasi-thermal-equilibrium distributions for hot electrons and holes in their subbands with very different temperatures.
Physically, the elapsed time for reaching such a quasi-thermal-equilibrium state is termed as an energy-relaxation time which depends on incident laser-pulse's width, intensity and excess energy $\hbar\omega_0-\varepsilon_{\rm G}$,
semiconductor band structure, lattice temperature and other material parameters.
As displayed in Fig.\,\ref{density}, the photo-generated carrier density decreases with reducing excess energy $\hbar\omega_0-\varepsilon_{\rm G}$ ($\varepsilon_{\rm G}=1.54\,$eV) due to down-shifts of carrier Fermi energies.

\medskip

The plotted $n_{j,{\rm 1D}}(t)$ in Fig.\,\ref{density} only reveals the change in the sum of occupations over all $k$ values as a function of time.
In order to visualize the time-dependent distribution of carriers in $k$ space, we can display the time-resolved photoluminescence spectra.
Having calculated the expression in Eq.\,\eqref{eq:pl}, we display in Fig.\,\ref{luminescence}
the time-evolution of photoluminescence spectra ${\cal P}_{j,{\rm pl}}(\Omega_0\,\vert\, t)$ resulting from electron-hole pair spontaneous recombinations within
the central quantum wire as the light pulse passes through the quantum wire, where $\hbar\Omega_0$ is the energy of spontaneously emitted photons. From this figure, we observe a sharp peak at the bandgap energy
$\hbar\Omega_0\approx\varepsilon_{\rm G}$ due to the presence of a very large peak at $k=0$ in the product of occupation factors $n_{j,k}^{\rm e}(t)n_{j,k}^{\rm h}(t)$ in Eq.\,\eqref{eq:pl}.
This photoluminescence peak is closely followed by an exponential-like long tail which results from spontaneous emission at $|k|>0$ electronic states and is determined by the line-shape function
$\sim n_{j,k}^{\rm e}(t)n_{j,k}^{\rm h}(t)\,L(\hbar\Omega_0-\varepsilon_{\rm G}-\varepsilon^{\rm e}_k
-\varepsilon^{\rm h}_k-\varepsilon_{j,{\rm c}}(k,\,t), \; \hbar\gamma_{\rm eh})$
with a negative time-dependent slope for hot carriers in the central quantum wire.
Here, it is very interesting to note that a phonon-hole burning appears as a cusp between two different slopes in the photoluminescence spectra around $\hbar\Omega_0-\varepsilon_{\rm G}\approx 65\,$meV
due to its dependence on $n_{j,k}^{\rm e}(t)$ in Eq.\,\eqref{eq:pl}.
The larger slope on the left-hand side of this cusp comes from the non-thermal population of electrons at low kinetic energies, while the smaller slope on the right-hand side of the cusp is attributed to the quasi-thermal-equilibrium population of electrons at high energies (tails beyond the phonon-hole burning in Fig.\,\ref{occupation}(a)).
This cusp from phonon-hole burning is gradually smoothened with time ($t=1.6\,$ps) and will be eventually filled up for $t\geq 10\,$ps (not shown) by nearby distributed electrons in $k$ space.
Moreover, the merging of slopes at different times in the range of $\hbar\Omega_0>1.78\,$eV reflects the dynamics of electrons and holes in their individual quasi-thermal-equilibrium states ($t\in 140\,$fs$-1.6\,$ps)
as shown in Fig.\,\ref{occupation}.

\begin{figure*}%[p]
\centering
\includegraphics[width=0.8\textwidth]{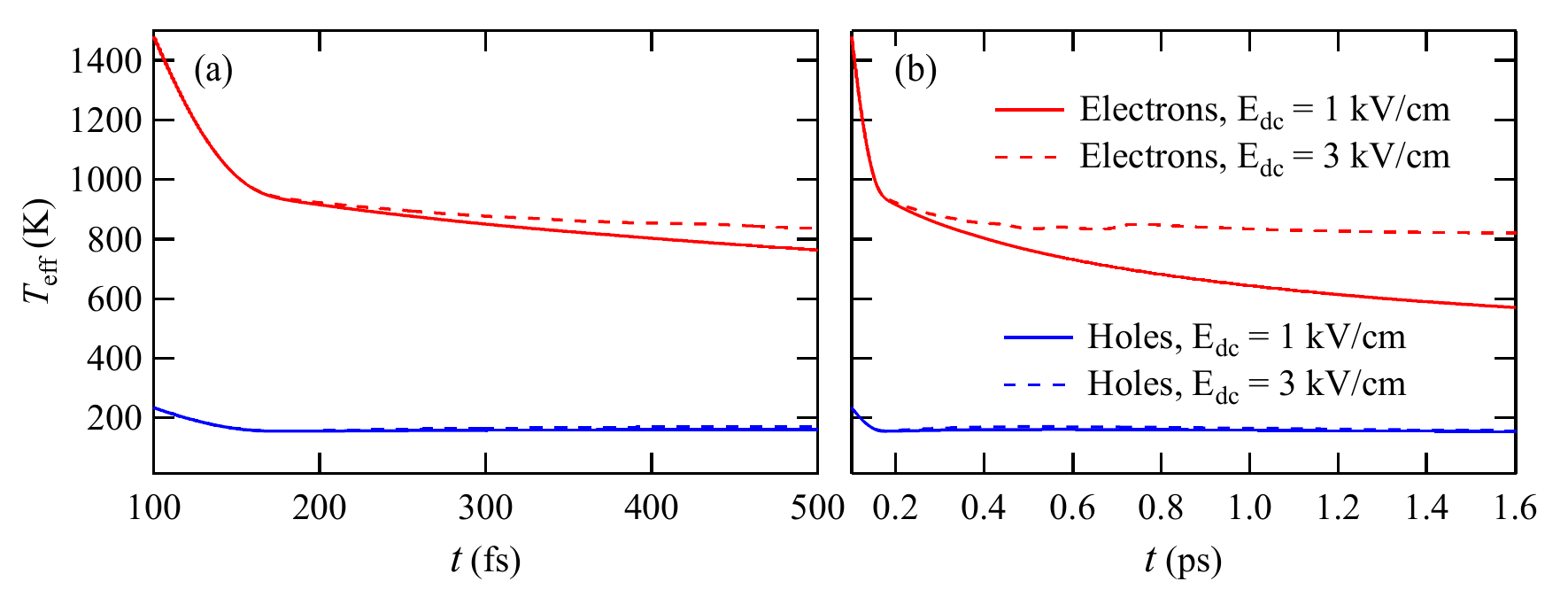}
\caption{(Color online) Calculated effective temperatures for electrons $T_{j,{\rm e}}(t)$ (red) and holes $T_{j,{\rm h}}(t)$ (blue) [see their expressions right after Eq.\,(\ref{eq:thengy})] as functions of $t$ within the central quantum wire for two values of DC electric field $E_{\rm dc}=1$ (solid) and $3\,$kV/cm (dashed).
Both the expanded ($a$) and the complete ($b$) views are presented. Here, a
$40\,$fs light pulse interacts significantly with quantum wires within the time interval $t\in100-180\,$fs.}
\label{temperature}
\end{figure*}

\medskip

For a non-thermal carrier distribution, no temperature can be defined physically for describing the thermodynamics of photo-excited carriers until a quasi-thermal-equilibrium state has been reached.
In this case, however, one can still define the so-called ``effective'' carrier temperature through a quantum-statistical average for kinetic energies of all these non-thermal carriers.
Based on calculated average kinetic energies ${\cal Q}_j^{\rm e,h}(t)$ from Eq.\,\eqref{eq:thengy} (not shown), the individual ``effective'' temperatures for electrons $T_{j,{\rm e}}(t)$ and holes $T_{j,{\rm h}}(t)$ can be obtained.
We present in Fig.\,\ref{temperature} the calculated $T_{j,{\rm e}}(t)$ (red solid) and $T_{j,{\rm h}}(t)$ (blue solid) of the central quantum wire at $E_{\rm dc}=1\,$kV/cm as functions of time, where both the short-time-scale (a) and the
long-time-scale (b) views are provided. As shown in Fig.\,\ref{temperature}(a), right after the front of the light pulse hits the quantum wire ($t\approx 100\,$fs), the photo-excited electrons become extremely hot in this non-thermal stage
with $T_{j,{\rm e}}(t)$ running as high as $\sim 1500\,$K ($77\,$K for the lattice temperature). This non-thermal stage for electrons extends all the way
until an initial thermalization time $t_0\sim 670\,$fs is reached, where $t_0$ can be determined from the variation of distributions $n^{\rm e}_{j,k}(t)$ with time in Fig.\,\ref{occupation}(a).
During this short period of time, $T_{j,{\rm e}}(t)$ quickly drops from $\sim 1500\,$K to $\sim 900\,$K through emission of many optical phonons,
as shown in Fig.\,\ref{temperature}(b). After the initiation of an electron thermal stage ($t>t_0$), $T_{j,{\rm e}}(t)$ only slowly decreases to $\sim 600\,$K at $t=1.6\,$ps due to electron-hole Coulomb scattering and continued phonon emmision.
On the other hand, $T_{j,{\rm h}}(t)$ drops very slowly from its initial value $\sim 250\,$K ($t\approx 100\,$fs) to $\sim 150\,$K at its initial thermalization time $t_1\sim 335\,$fs and remains nearly constant thereafter,
where $t_1$ can also be estimated from the change of distributions $n^{\rm h}_{j,k}(t)$ with time in Fig.\,\ref{occupation}(b). Throughout this overall cooling process, ``cool'' holes are heated by hot
electrons through electron-hole Coulomb scattering and $T_{j,{\rm h}}(t)$ changes from decreasing to increasing with time after $t=200\,$fs.
\medskip

In addition to heating carriers with a laser pulse, an applied DC electric field $E_{\rm dc}$ can also heat carriers through a resistive force acting on field-driven carriers, i.e., Joule (or Ohmic) heating.
Such a Joule heating is expected to increase quadratically with $E_{\rm dc}$, especially within the nonlinear-transport regime under a high DC field.
For a stronger DC electric field $E_{\rm dc}=3\,$kV/cm, from Fig.\,\ref{temperature} we find Joule heating starts taking over reducing laser heating of electrons around $t\sim 250\,$fs (red dashed), and $T_{j,{\rm e}}(t)$ is sustained at $\sim 800\,$K thereafter,
instead of a dropping $T_{j,{\rm e}}(t)$ with time under a lower DC field $E_{\rm dc}=1\,$kV/cm (red solid).
Since the resistive force acting on holes is much smaller due to their slow drifting motions, Joule-heating effect on them becomes insignificant and there is no visible change in the results of $T_{j,{\rm h}}(t)$
(blue solid and dashed) for $E_{\rm dc}=1$ and $3\,$kV/cm within the nonlinear-transport regime of photo-excited carriers.

\begin{figure*}%[p]
\centering
\includegraphics[width=0.8\textwidth]{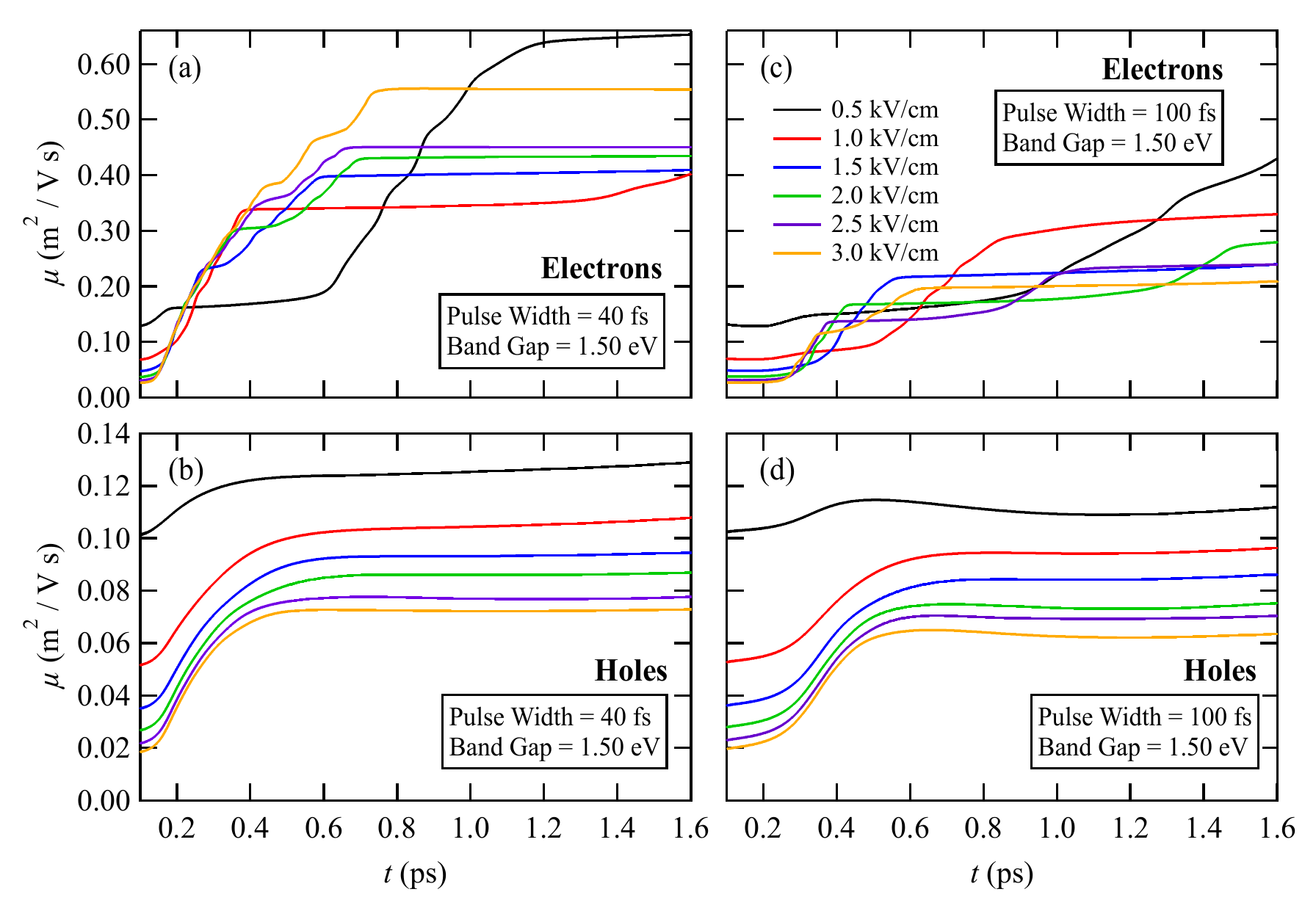}
\caption{(Color online) Calculated mobilities for electrons $|\mu_j^{\rm e}(t)|$ [($a$),($c$)] and holes $\mu_j^{\rm h}(t)$ [($b$),($d$)] from Eq.\,(\ref{eq:vg}) as functions of $t$ within the central quantum wire with six values of DC electric field
$E_{\rm dc}$ from $0.5$ to $3.0\,$kV/cm in steps of $0.5\,$kV/cm.
Both results for a $40\,$fs [($a$),($b$)] and a $100\,$fs [($c$),($d$)] pulse are presented. Here, the light pulse interacts significantly with quantum wires within the time interval $t\in100-180\,$fs.}
\label{mobility}
\end{figure*}

\medskip

After the generation of photo-excited electron-hole pairs in quantum wires by incident light pulse, these carriers are driven by an applied DC electric field $E_{\rm dc}$, leading to asymmetric distributions
$n_{j,k}^{\rm e,h}(t)$ with respect to $k=0$ and nonzero drift velocities $v_j^{\rm e}(t)$ for electrons and $v_j^{\rm h}(t)$ for holes.
The transient $v_j^{\rm e}(t)$ and $v_j^{\rm h}(t)$ can be calculated by using Eq.\,\eqref{eq:vg} as statistical-averaged group velocities of electrons and holes respectively.
The resistive forces, given by the second terms in Eqs.\,\eqref{eq:Fe} and \,\eqref{eq:Fh} for electrons and holes, are the reason for Joule heating of these photo-excited carriers under a strong $E_{\rm dc}$.
Such a heating process is directly connected to momentum dissipation of driven carriers, which leads to a saturation of carrier drift velocities under a strong DC field.
In Fig.\,\ref{mobility} we show the calculated electron and hole mobilities ($\mu_j^{\rm e,h}(t) = |v_j^{\rm e,h}(t)| / E_\mathrm{dc}$) in the central quantum wire as functions of time $t$.
Plots are shown for these cases of exposure to a 40 fs pulse (a,b) as well as a 100 fs pulse (c,d) of the same total energy. Each plot presents simulation results using a different DC electric field applied to the wire.
\medskip

In the linear-transport regime, $\mu_j^{\rm e,h}(t)$ should be independent of $E_{\rm dc}$ although they may still vary with time due to transient occupations $n_{j,k}^{\rm e,h}(t)$ produced by a laser pulse.
For a strong $E_{\rm dc}$, however, nonlinear transport of these photo-excited carriers occurs, leading to decreasing $\mu_j^{\rm h}(t)$ with $E_{\rm dc}$, as shown in Figs.\,\ref{mobility}(b) and \ref{mobility}(d)
for a quasi-thermal-equilibrium distribution ($t=1.6\,$ps) of photo-generated holes.
For non-thermal photo-generated electrons, on the other hand, we find $n_{j,k}^{\rm e}(t)$, as a function of $k$, changes dramatically with time in Fig.\,\ref{occupation}(a).
This leads to a large drop of $\mu_j^{\rm e}(t)$ with increasing $E_{\rm dc}$ from $0.5\,$kV/cm to $1\,$kV/cm due to Joule heating,
which is followed by a gradual increase of $\mu_j^{\rm e}(t)$ with $E_{\rm dc}$ from $1\,$kV/cm up to $3\,$kV/cm, as presented in Fig.\,\ref{mobility}(a).
The enhancement of $\mu_j^{\rm e}(t)$ with $E_{\rm dc}$ results from a DC-field induced Doppler shift in both absorption and emission of longitudinal-optical phonons, as demonstrated by Eqs.\,\eqref{eq:eem} and \eqref{eq:eabs}.
These changes in phonon absorption and emission will affect energy relaxation of hot electrons ($t>t_0\sim 670\,$fs), modifying time dependence of $\mu_j^{\rm e}(t)$ in Fig.\,\ref{mobility}(a) with various $E_{\rm dc}$.
However, such a Doppler-shift effect becomes negligible for holes due to their much smaller drift velocity compared to that of electrons.
Additionally, since a longer pulse can cause major modification to the non-thermal distribution of photo-excited electrons in $k$ space with time, we expect different time evolutions of $\mu_j^{\rm e}(t)$
with various $E_{\rm dc}$, as displayed in Fig.\,\ref{mobility}(b).

%%%%%%%%%%%%%%%%%%%%%%%%%%%%%%%%%%%%%%%%%
\begin{figure*}%[p]
\centering
\includegraphics[width=0.8\textwidth]{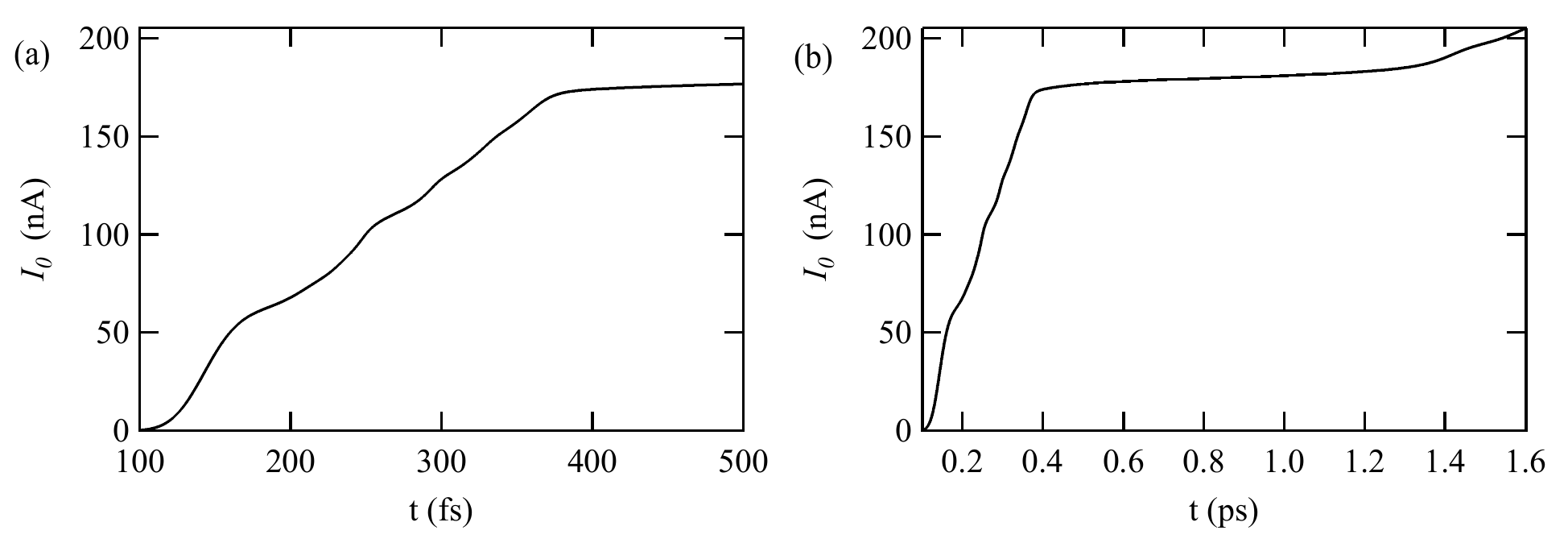}
\caption{Calculated photo-current $I_{j,{\rm ph}}(t)=J_{j,{\rm ph}}(t)(2\delta_0/\alpha)$ from Eq.\,(\ref{eq:dcj}) at $E_{\rm dc}=1\,$kV/cm as a function of $t$ within the central quantum wire.
The expanded ($a$) and the complete ($b$) views are presented.
Here, a $40\,$fs light pulse interacts significantly with quantum wires within the time interval $t\in100-180\,$fs.}
\label{current}
\end{figure*}

\medskip

Besides the time-resolved photoluminescence spectra in Fig.\,\ref{luminescence}, another direct measurement for studying transient electronic properties comes from photo-current
which involves both transient charge density induced by a laser pulse and transient drift velocity driven by a DC electric field.
The transient induced photo-current $I_{j,{\rm ph}}(t)$ determined from Eq.\,\eqref{eq:dcj} in the central quantum wire is exhibited in Fig.\,\ref{current}, where both the short-time-scale (a) and the
long-time-scale (b) views are provided.
From Fig.\,\ref{current}(a), we find $I_{j,{\rm ph}}(t)$ initially increases very rapidly with $t$ as the pulse maximum is entering into the quantum wire ($100\leq t\leq 200\,$fs).
This observed behavior is related to the fact that $n^{\rm e,h}_{j,{\rm 1D}}(t)$ are being built up very fast during this fast-increasing period of time, as shown in Fig.\,\ref{density}.
After this initial short period of time, the increasing rate of $I_{j,{\rm ph}}(t)$ slightly decreases for the slow-increasing period of time ($200\leq t\leq 400\,$fs), where the linear density $n_{j,{\rm 1D}}(t)$ is already independent of $t$
but drift velocities $v_j^{\rm e,h}(t)$ still linearly increase with $t$ approximately (not shown). As $t$ goes beyond $400\,$fs up to $1.6\,$ps,  $I_{j,{\rm ph}}(t)$ becomes nearly a constant, as seen from Fig.\,\ref{current}(b),
where both $n_{j,{\rm 1D}}(t)$ and $v_j^{\rm e,h}(t)$ become time independent and a much longer carrier-cooling process, as shown in Fig.\,\ref{temperature}, starts.
Technically, if the slow-increasing period ($200\,\leq t\leq 400\,$fs) can be shortened and the saturated photo-current ($400\,$fs$\leq t\leq 1.6\,$ps) can be eliminated at the same time with high extrinsic defects,\,\cite{THW}
the first fast-increasing part in $I_{j,{\rm ph}}(t)$ can possibly be used for the generation of a THz-wave. Here,
the fast and slow increasing periods of time correspond, respectively, to the quantum kinetics of $n^{\rm e,h}_{j,{\rm 1D}}(t)$, due to incident femtosecond light pulse,
and to the thermal dynamics of $v_j^{\rm e,h}(t)$, associated with reshaping $n^{\rm e,h}_{j,{\rm 1D}}(t)$ into a quasi-thermal-equilibrium distribution.

\subsection{Transient Light-Field and Light-Wire Interaction Properties}
\label{disc-2}

In addition to local measurements of both photo-current and photoluminescence spectra, we can also detect changes in propagating laser pulses far away from quantum wires to explore further the interaction dynamics of a transient optical field.
Specifically, we would like to address how a propagating electric-field component of a laser pulse is affected by a locally-induced optical-polarization field as a back action of electrons in quantum wires on interacting laser photons, and vice versa.
Such an electron back action will contain both transverse dipole-induced and a longitudinal plasma-wave-induced polarization fields, as elucidated by Eqs.\,\eqref{eq:qwpol} and \eqref{eq:Pq} for the former and by Eqs.\,\eqref{eq:rhoh} and \eqref{eq:rhoe} for the latter.

\begin{figure*}%[p]
\centering
\includegraphics[width=0.8\textwidth]{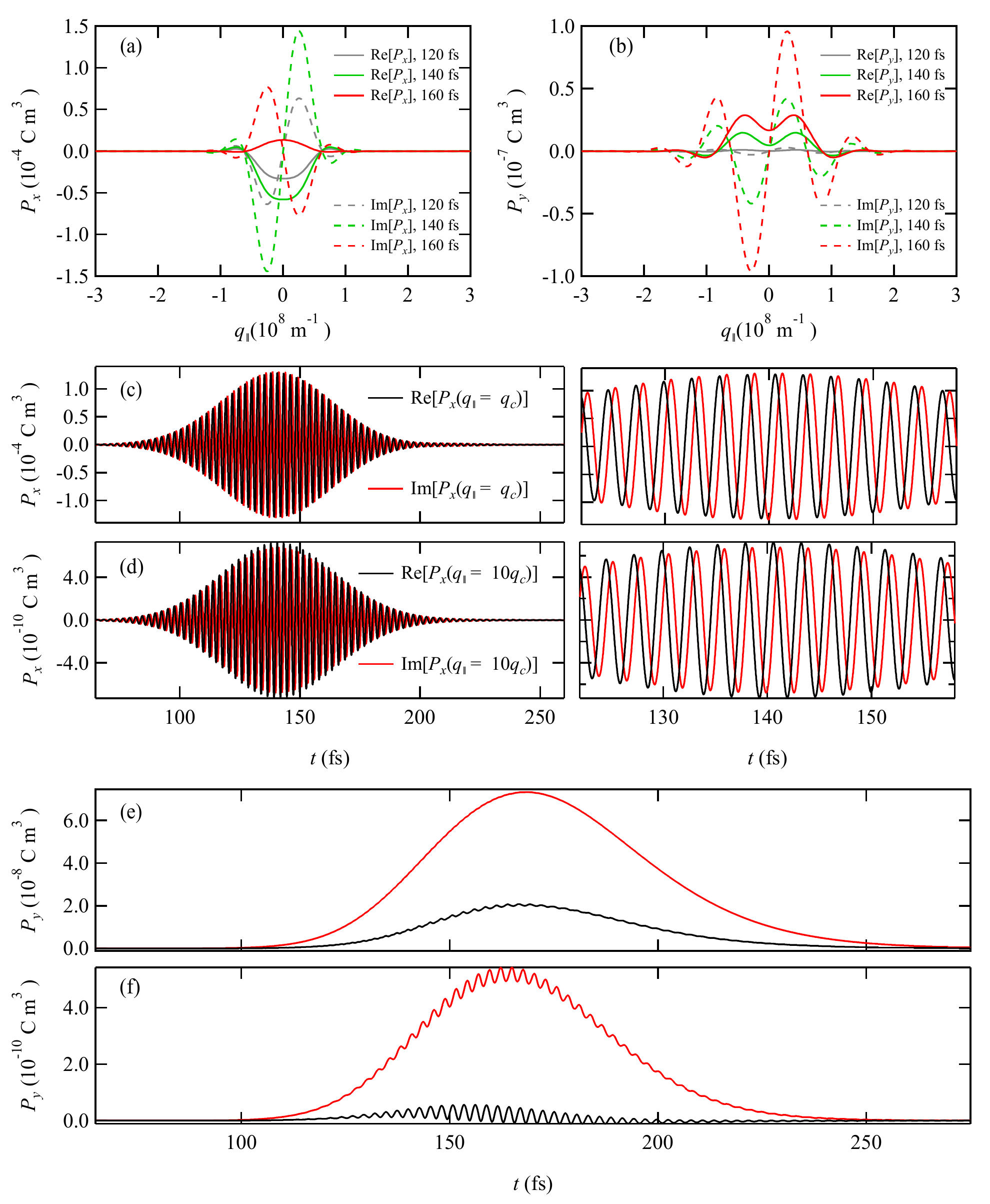}
\caption{(Color online) Calculated real and imaginary parts of $\tilde{P}_j^x(q_\|,t)$ [(a) (c),(d)] and $\tilde{P}_j^y(q_\|,t)$ [(b),(e),(f)] from Eq.\,(\ref{eq:Pq})
as a function of wave number $q_\|$ [(a), (b)] (real-solid, imaginary-dashed) at different times
$t=120,\,140,\,160\,$fs within the central quantum wire as well as a function of time $t$ [(c)-(f)] (real-black, imaginary-red) at $q_\|/q_0=1$ [(c),(e)]
and $10$ [(d),(f)]. Here, the light pulse interacts significantly with quantum wires within the time interval $t\in100-180\,$fs.}
\label{polarization}
\end{figure*}

\medskip

As a starting point, we first show how a quantum-kinetic (microscopic) optical coherence\,\cite{od} is self-consistently established by photo-excited electron-hole pairs as an optical response to a total electric field including its own generated polarization field.
The 1D quantum-wire polarization components $\tilde{P}^{x,y}_j(q_\|,t)$ from Eq.\eqref{eq:Pq} in $q_\|$ space are presented in Figs.\,\ref{polarization}(a) and \ref{polarization}(b), and also as functions of time at both small and large values of $q_\|$ in Figs.\,\ref{polarization}(c)$-$\ref{polarization}(f).
$P^{x,y}_j(q_\|,t)$ are complex in $q_\|$ space, so both the real and imaginary parts of them are displayed in these four panels.
Since the polarization in $y$-space must be real, we see that the real and imaginary parts in Figs.\,\ref{polarization}(a) and \ref{polarization}(b) are symmetric and antisymmetric about $q_\|=0$, respectively.
Moreover, the dipole polarization field in Fig.\,\ref{polarization}(a) becomes much stronger than the plasma-wave polarization field in Fig.\,\ref{polarization}(b) due to the dominant $x$-polarized electric-field component in the incident laser pulse.
Furthermore, the $q_\|$-space spreading of the plasma-wave polarization field is found to be broader than that of the dipole polarization field, implying a stronger localization in the $x$ direction for the former.
\medskip

For time dependence, Figs.\ \ref{polarization}(c) and \ref{polarization}(d) indicate that the dipole polarization field oscillates rapidly at the laser-field frequency $\omega_0$, as expected since the laser field is polarized in the $x$ direction.
However, at the array center the polarization field in the $y$ direction, resulting from the longitudinal plasma waves created by light induced charge-density fluctuations along the wire, becomes the only nonzero one.
Figures\ \ref{polarization}(e) and \ref{polarization}(f) further reveal that this plasma-wave polarization field does not oscillate with the laser frequency.
Instead, it has the time dependence of the temporal laser envelope, but with an extended tail in time, reflecting the existence of charge-density fluctuations only during the time period for the presence of a pulsed laser within the wire.
The Fourier transform with respect to time of Fig.\,\ref{polarization}(e) results in a distribution centered about $\omega=0$, and with a width of about $20\,$THz.
This implies a very low plasmon energy (very slow oscillations with a very long time period) for such a small $q_\|$ value used in Fig.\,\ref{polarization}(e),
but the plasmon energy increases greatly at a much large $q_\|$ value in Fig.\,\ref{polarization}(f),
i.e., fast oscillations with a much shorter time period.
One possible application of this work is to determine what combination of pulses, when sent through a quantum-wire array, will generate a localized plasma-wave polarization field with a desired THz spectrum\,\cite{THW,xczhang},
which can then be transformed into a propagating transverse electric field after employing a surface grating.

\begin{figure*}%[p]
\centering
\includegraphics[width=0.7\textwidth]{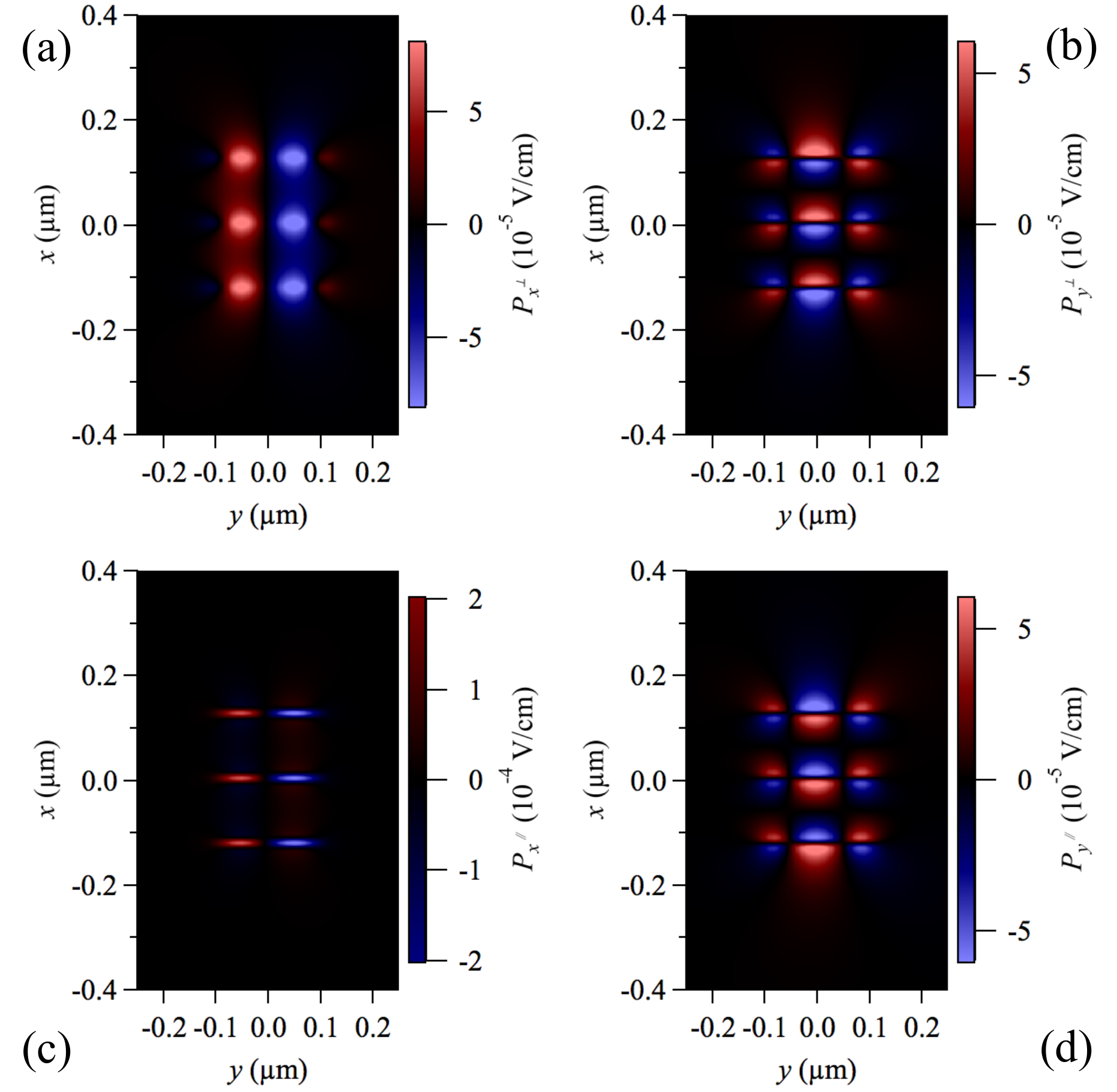}
\caption{(Color online) Density plots for spatial distributions of both transverse [(a),(b)] and longitudinal [(c),(d)] local polarization fields at the moment of $t=140\,$fs
that a $40\,$fs pulse peak is passing through the middle of a quantum-wire array ($y=0$),
where transverse- and longitudinal-polarization-field components along both the $x$ [(a),(c)] and $y$ [(b),(d)] directions are displayed.
The linear array consists of three quantum wires displaced in the $x$ direction at $x=0$ and $\pm 125\,$nm, respectively.}
\label{pfield-2}
\end{figure*}

\medskip

The back action\,\cite{backaction} from photo-excited electron-hole pairs in quantum wires on incident laser photons can be analyzed by studying self-consistent macroscopic optical polarization fields
generated by induced dipole moments (plasma-waves) perpendicular (parallel) to the wires.
The plots for the localized transverse ($P_{x,y}^\perp$) and longitudinal ($P_{x,y}^\|$) quantum-wire polarizations are displayed in Fig.\,\ref{pfield-2}
in a region near the wire array.
Here, the array is centered about $x=0$ and $y=0$.  These three wires are separated by $a=125\,$nm along the $x$-direction.
Quantitatively, we note that all the polarization contributions, $P_x^\perp(x,y)$ and $P_y^\perp(x,y)$, are comparable in strength.
The strongest polarization contribution, $P_x^\|(x,y)$, results from the strong bound charge density and varies rapidly around the vicinity of the wires.
The diffraction of the incident laser pulse by this small array is significant, and then the higher-order diffracted light beam, which acquires a very
large angle with respect to the $y$ direction (or $q_\perp\gg q_\|$), gives rise to a very strong longitudinal polarization field $P_x^\|(x,y)$
in Fig.\,\ref{pfield-2}(c).
We also find similarities in the spatial distributions for $P_x^{\|,\perp}(x,y)$ in Figs.\,\ref{pfield-2}(a) and \ref{pfield-2}(c)
as well as for $P_y^{\|,\perp}(x,y)$ in Figs.\,\ref{pfield-2}(b) and \ref{pfield-2}(d). These local electronic ``fingerprints'' in $P_x^{\perp}(\mbox{\boldmath$r$},t)$
can still be embedded in and further carried away by a propagating transverse electric field $E_x^\perp(\mbox{\boldmath$r$},t)$ over a very large distance.
Here, the phases of both $P_x^{\|,\perp}(x,y)$ and
$P_y^{\|,\perp}(x,y)$ in each quantum wire remain the same due to the lack of inter-wire electromagnetic (Coulomb) coupling for the large wire separation ($a=125\,$nm).
Only $P_x^{\perp}(x,y)$ in Fig.\,\ref{pfield-2}(a) becomes delocalized within the wire array along the $x$ direction, while the other three in Figs.\,\ref{pfield-2}(b)$-$\ref{pfield-2}(d) are kept localized.
The distributions above and below a wire in Figs.\,\ref{pfield-2}(b) and \ref{pfield-2}(d) acquire opposite phases, while 
the distributions of each wire in Figs.\,\ref{pfield-2}(a) and \ref{pfield-2}(c) keeps the same phase.    
Furthermore, the similarity between  Figs.\,\ref{pfield-2}(b) and \ref{pfield-2}(d) indicates that the electronic fingerprint from the local e-h plasma waves 
can be imprinted on the e-h pair dipole moments, and vice versa as shown in Figs.\,\ref{pfield-2}(a) and \ref{pfield-2}(c).

\begin{figure*}%[p]
\centering
\includegraphics[width=0.7\textwidth]{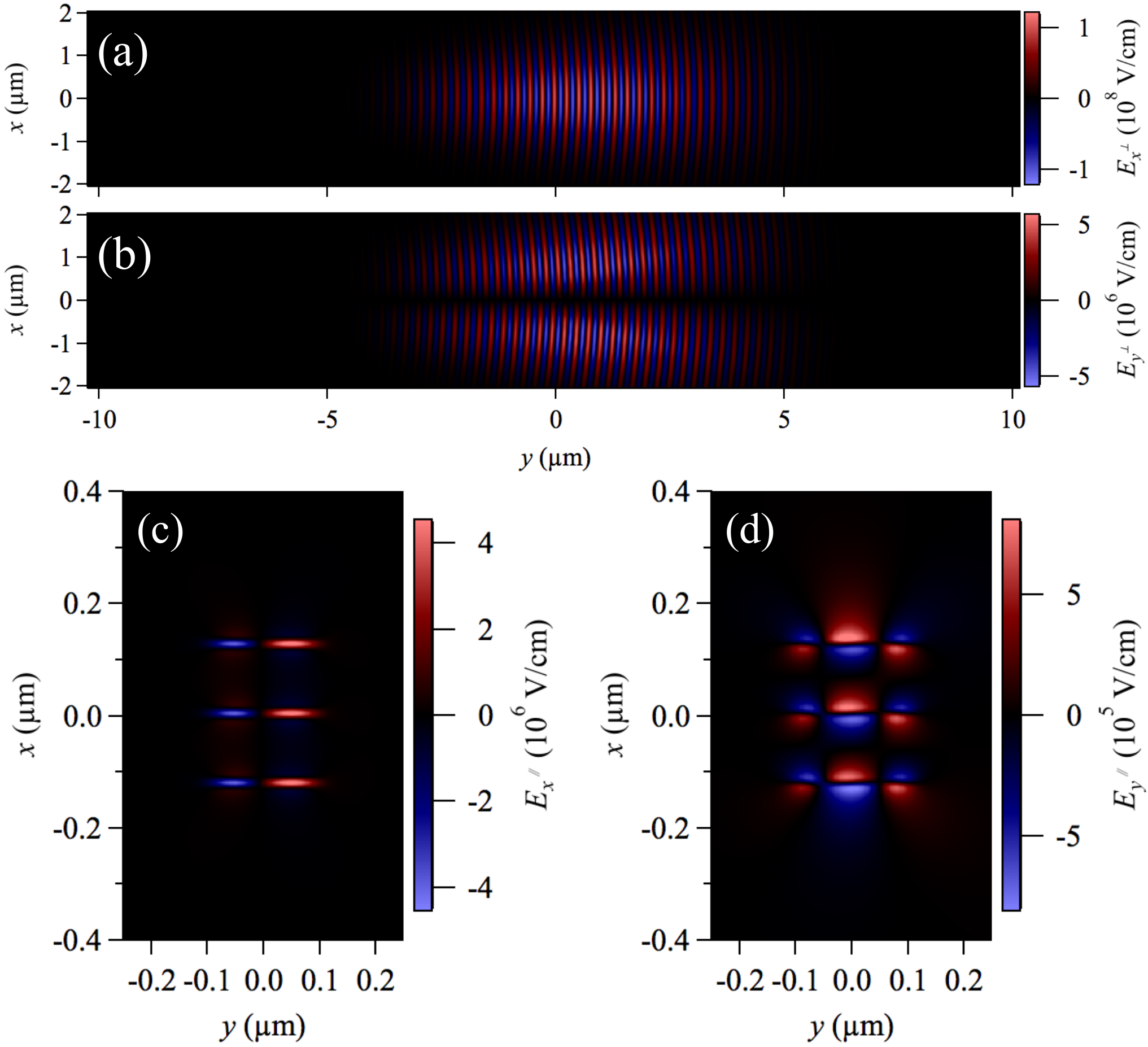}
\caption{(Color online) Density plots for spatial distributions of both transverse [(a),(b)] and longitudinal [(c),(d)] propagating electric fields at the moment of $t=140\,$fs
that a $40\,$fs pulse peak is passing through the middle of a quantum-wire array ($y=0$),
where transverse- and longitudinal-field components along both the $x$ [(a),(c)] and $y$ [(b),(d)] directions are presented.
The array used is the same as that in Fig.\,\ref{pfield-2}.}
\label{efield-2}
\end{figure*}

\medskip

In order to show the effect of back action from photo-excited electron-hole pairs in quantum wires on incident laser photons, we need to study the dynamics of both
propagating transverse and localized longitudinal electric-field components.
Figure\ \ref{efield-2} displays the propagating transverse ($E_{x,y}^\perp$) and localized longitudinal ($E_{x,y}^\|$) electric fields
as the center of the $40\,$fs laser pulse passes through the same three-wire array.  Note that the longitudinal electric fields are shown only around the vicinity of the wires. Note also the difference in scales between the plots in Fig.\,\ref{efield-2}.
For example, the maximum field strength for $E_x^\perp(x,y)$ in Fig.\,\ref{efield-2}(a) is $1.25\,$MV/cm, whereas the maximum field strength for $E_y^\perp(x,y)$ in Fig.\,\ref{efield-2}(b) is only $0.05\,$MV/cm.
This is because the laser pulse is primarily polarized in the $x$ direction, but the tight focusing conditions create a small but significant transverse $y$-component.
Also notable is that the peak magnitude of $E_x^\|(x,y)$ is an order of magnitude greater than that of $E_y^\|(x,y)$,
a fact that only a multi-dimensional propagation model in this paper will reveal.
The fact that $E_x^\|(x,y)$ is an order of magnitude bigger than $E_y^\|(x,y)$ in Fig.\,\ref{efield-2} can be explained in the same way as the
occurrence of the strongest $P_x^\|(x,y)$ in Fig.\,\ref{pfield-2}(c).
However, as shown below, this reasoning does not hold for larger arrays with smaller inter-wire spacing.
The comparison of Figs.\,\ref{efield-2}(a) and \ref{efield-2}(b) clearly demonstrates that the diffraction of the laser pulse only appears for $E_y^\perp(x,y)$ but not for $E_x^\perp(x,y)$. Moreover, $E_x^\|(x,y)$ in Fig.\,\ref{efield-2}(c) and
$E_y^\|(x,y)$ in Fig.\,\ref{efield-2}(d) are reminiscent of the corresponding features of $P_x^\|(x,y)$ and $P_y^\|(x,y)$, respectively,
in Figs.\,\ref{pfield-2}(c) and \ref{pfield-2}(d) but with an opposite phase as can be verified by Eq.\,\eqref{eq:E}.
Therefore, these imprinted electronic fingerprints on the local polarization fields $P_{x,y}^{\|,\perp}(\mbox{\boldmath$r$},t)$ can be transferred from quantum wires to a distant place by propagating 
electric-field components $E_{x,y}^{\|,\perp}(\mbox{\boldmath$r$},t)$, respectively, if
a conversion from longitudinal to transverse electric field can be fulfilled using a surface grating.
Another possible application of the current work is to make use of the correlation between the local $P_{x,y}^{\perp}(\mbox{\boldmath$r$},t)$ fields and the
remote $E_{x,y}^{\perp}(\mbox{\boldmath$r$},t)$ fields for extraction of a photon quantum memory\,\cite{op-memory} in the far-field region.

\begin{figure*}%[p]
\centering
\includegraphics[width=0.8\textwidth]{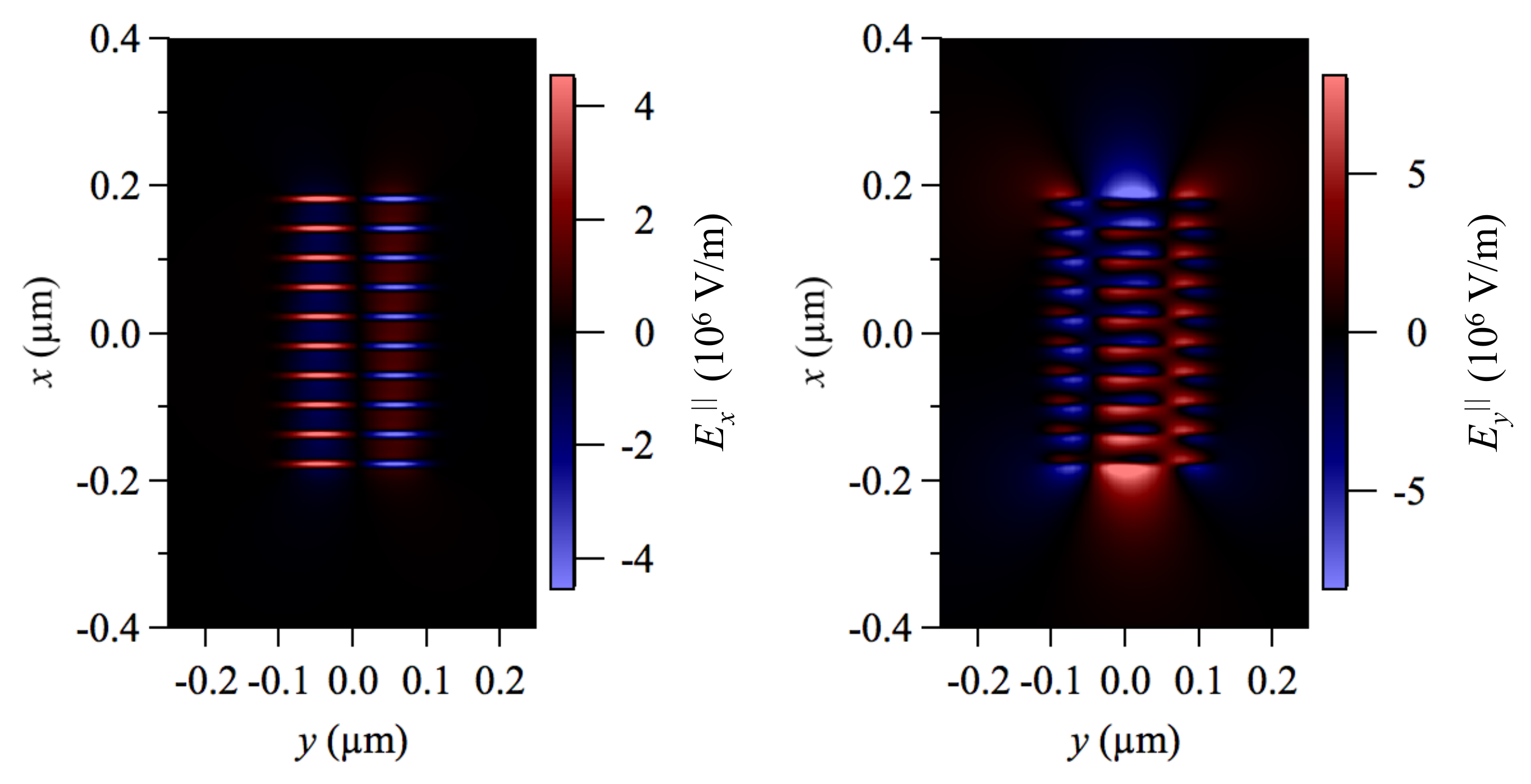}
\caption{(Color online) Density plots for spatial distributions of the longitudinal electric fields at the moment of $t=140\,$fs
that a $40\,$fs pulse peak is passing through the middle of a quantum-wire array,
where longitudinal-field components along both the $x$ [(a)] and $y$ [(b)] directions are presented. The linear array consists of ten quantum wires displaced in the $x$ direction at intervals of $40\,$nm.}
\label{2d_long_10wires}
\end{figure*}
\begin{figure*}%[p]
\centering
\includegraphics[width=0.8\textwidth]{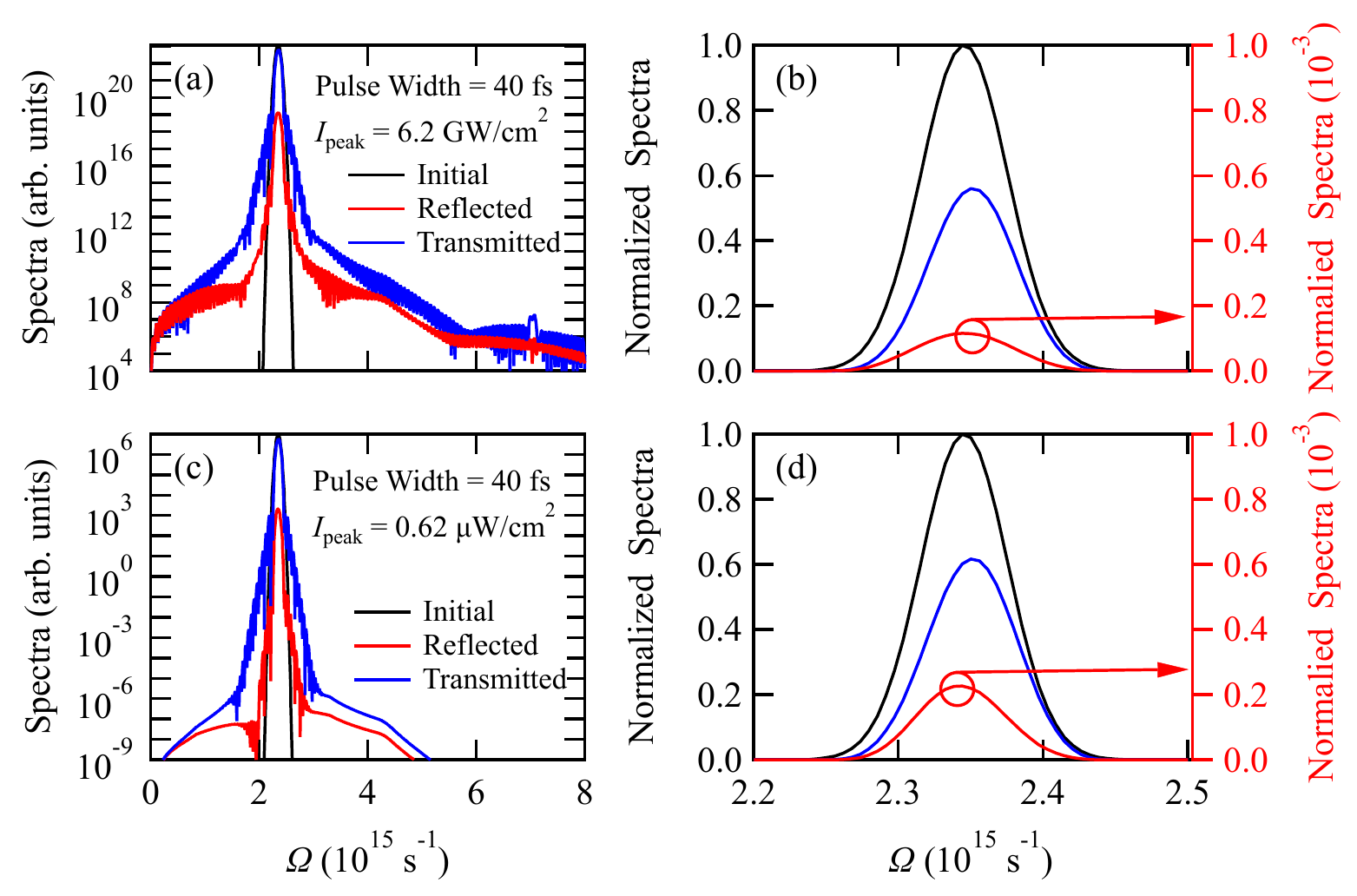}
\caption{(Color online) Calculated intensity ratios for incident (black), reflected (red) and transmitted (blue) in both logarithm [($a$),($c$)] and linear [($b$),($d$)] scales for $40\,$fs light pulses with
peak intensities $6.2\,$GW/cm$^{2}$  [($a$),($b$)] and $0.62\,$kW/cm$^{2}$ [($c$),($d$)]
from Eqs.\,\eqref{trans} and \eqref{refle} as functions of Fourier frequency $\Omega$.}
\label{fourier_spec}
\end{figure*}

\medskip

From the discussions of Fig.\,\ref{efield-2}, we know that both the laser-pulse diffraction by a wire array and the inter-wire Coulomb coupling can
play an important role in spatial distributions of $\mbox{\boldmath$E$}^{\perp,\|}(x,y)$ around the vicinity of the wire array.
Figure\ \ref{2d_long_10wires} presents $E_{x,y}^\|(x,y)$ as the center of the $40\,$fs laser pulse passes through a ten-wire array
[$E_{x,y}^\perp(x,y)$ are the same as in Fig.\,\ref{efield-2}(a) and \ref{efield-2}(b)].
Again, the array is centered about $x=0$ and $y=0$, but the ten wires are each separated by $a=40\,$nm along the $x$ direction.
$E_x^\|(x,y)$ in Fig.\,\ref{2d_long_10wires}(a) appears much as one might expect when comparing to Fig.\,\ref{efield-2}(c),
but we note that $E_y^\|(x,y)$ is now stronger than $E_x^\|(x,y)$.
This is because the smaller spacing between the wires leads to mutual interactions between electrons in different wires,
causing a strong nonlocal electro-optical interaction\,\cite{book-huang} between the wires.
Additionally, the array structure in Fig.\,\ref{2d_long_10wires}(b) is not as clear as the other field profiles.
This is due to the structure and diffraction of the small, but significant, $E_y^\perp(x,y)$ laser field component in Fig.\,\ref{efield-2}(b).
The $E_y^\perp(x,y)$ is zero at $x=0$ (between the two central wires),
but gets stronger on the edges of the array, increasing the impact on the nonlinear optoelectronic response\,\cite{nlo} 
for the outer wires. Although the phases of $E_x^\|(x,y)$ in Fig.\,\ref{2d_long_10wires}(a) for each wire still stay the same, the phases of $E_y^\|(x,y)$, 
associated with each wire in Fig.\,\ref{2d_long_10wires}(b), change from $q_\perp a=0$ to $q_\perp a=\pi$ between the top and bottom wires.
In the presence of strong inter-wire Coulomb coupling for a ten-wire array, the single-wire plasmon mode is split into ten different ones\,\cite{wirearray1,wirearray2,wirearray3},
having the highest energy for the in-phase mode ($q_\perp a=0$) down to the lowest energy for the out-of-phase mode ($q_\perp a=\pi$).
The inter-wire Coulomb coupling scales with $\sim\exp(-|q_\||a)$ which increases with decreasing $q_\|$ and $a$ values\,\cite{book-huang}.
We also emphasize that $E_x^\|(x,y)$ in Fig.\,\ref{2d_long_10wires}(a) becomes delocalized for small spacing $a$, in contrast to the result in Fig.\,\ref{efield-2}(c) for large wire separation.
\medskip

Generally speaking, a linear-optical response of electrons to incident laser is independent of the laser-field strength. However, a nonlinear-optical response of electrons will decrease with increasing laser intensity.
In order to demonstrate nonlinear optoelectronic effects in our system, we present both transmission, $\mathbbmsl{T}_{{\rm F}}(\Omega\,\vert\,\omega_{0})$, and reflection, $\mathbbmsl{R}_{{\rm F}}(\Omega\,\vert\,\omega_{0})$, 
spectra from Eqs.\,\eqref{trans} and \eqref{refle} 
for the central quantum wire with a high peak intensity $I_{\rm peak}=6.2\,$GW/cm$^{2}$ in Fig.\,\ref{fourier_spec}(a) and a low peak intensity $0.62\,$kW/cm$^{2}$ in Fig.\,\ref{fourier_spec}(c).
As $I_{\rm peak}$ increases, we find the weak (strong) peak of normalized  $\mathbbmsl{R}_{{\rm F}}(\Omega\,\vert\,\omega_{0})$ [$\mathbbmsl{T}_{{\rm F}}(\Omega\,\vert\,\omega_{0})$] in Fig.\,\ref{fourier_spec}(d) 
is reduced and broadened simultaneously in Fig.\,\ref{fourier_spec}(b), as seen from Fig.\,\ref{fourier_spec}(a) for a much more clear view of broadening. 
The major peak reduction and broadening effects observed for $\mathbbmsl{R}_{{\rm F}}(\Omega\,\vert\,\omega_{0})$ are attributed to decreasing nonlinear optical response 
of the quantum wire to the intense incident laser field for the former, as well as to the enhanced optical dephasing rate with increasing $I_{\rm peak}$ for the latter.
Furthermore, the peak shift in $\mathbbmsl{R}_{{\rm F}}(\Omega\,\vert\,\omega_{0})$ is also observed for increasing $I_{\rm peak}$, which is connected to deformed fast oscillations within the wavepacket of reflected electromagnetic wave 
due to nonlinear dependence on laser field.  

\section{Conclusions and Remarks}
\label{conclu}

In this work we present a unified quantum-kinetic model for both optical excitations and transport of electrons in low dimensional solids within a single frame. The model is a self-consistent many-body theory for coupling the ultrafast carrier-plasma dynamics in a linear array of quantum wires with the scattering of ultrashort light pulses.
It couples the unified quantum-kinetic theory (beyond the perturbation approach) for the quantum wires self-consistently to Maxwell's equations for the field propagation without making any assumptions about the field structure (e.g., monochromatic, plain wave, purely transverse, uniform field in the quantum solid, etc.).
The quantum wire electron and hole distributions are evolved with optical excitations and many-body effects while being further driven by an applied DC electric field along the wires.
The many body-effects include collisions and resistive forces from intrinsic phonon and Coulomb scattering of the carriers.
This applied DC field significantly modifies the non-equilibrium properties of the induced electron-hole plasma, while the induced longitudinal electric fields of each wire contributes strongly to the nonlocal response from neighboring wires.
By including the longitudinal field effects on the wires and the resulting wire polarization, this model allows researchers to optimize the spectra and intensity of a radiated terahertz field by the photo-response of the quantum-wire array.
More generally, this model provides a quantitative tool for exploring ultrafast dynamics of pulsed light beams through quantum solids of reduced dimensionality.
Because the model itself makes no assumption of pulse parameters, it is ideal for calculating multi-frequency pulse correlations for single (different times) and dual (separated positions) light pulses to study the quantum kinetics of photo-excited electron-hole pairs.
It also serves as a basis model for the determination of multi-pulse damage thresholds for state-of-the-art nano-optoelectronic components.
\medskip

In this paper, we have addressed the following three fundamental physics issues using our model system in Fig.\,\ref{illustration}, i.e.,
(i) how the local transient photo-current and photoluminescence spectra are affected by laser pulse width, central frequency and intensity;
(ii) how the propagation of transverse and longitudinal electric-field components of a laser pulse are modified by an applied DC field;
(iii) how the stored local electronic fingerprints in self-consistently generated optical-polarization fields are carried away by incident laser pulse.
The corresponding applications of this research include determining the best combination of pulse sequence through a quantum-wire array to generate a localized plasma-wave polarization field with a desired THz spectrum\,\cite{THW,xczhang}, transferring local electronic fingerprints in polarization fields by a laser pulse
for the remote extraction of the stored photon quantum memory\,\cite{photonmem}, and ultra-fast optical modulations of nonlinear carrier transport by a laser pulse\,\cite{optotrans}.

\begin{acknowledgements}
JRG would like to acknowledge the financial supports from the Air Force Office of Scientific Research (AFOSR) through Contract No FA9550-13-1-0069
and also through Air Force Summer Faculty Fellowship Program (AF-SFFP).
DH thanks the supports from both the Air Force Office of Scientific Research (AFOSR) and the Laboratory University Collaboration Initiative (LUCI) program. 
\end{acknowledgements}

%\bibliography{jgulley}

\newpage
\appendix

\section{Density Distributions and Occupation Numbers}
\label{app1}

In second quantization, the electron number density operator in real space is
expressed as\,\cite{HaugKoch}
\begin{equation}
\label{rho}
\hat{\rho}_{\rm e}(y)
=
\hat{\psi}^\dagger_{\rm e}(y) \, \hat{\psi}_{\rm e}(y)\ ,
\:\:\:\:\:\:\:\:\:
\hat{\rho}_{\rm h}(y)
=
\hat{\psi}^\dagger_{\rm h}(y) \, \hat{\psi}_{\rm h}(y)\ ,
\end{equation}
where $\hat{\psi}^\dagger_{\rm e}(y)$ [$\hat{\psi}^\dagger_{\rm h}(y)$] and $\hat{\psi}_{\rm e}(y)$ [$\hat{\psi}_{\rm h}(y)$]
are creation and destruction field operators of electrons (holes) at the position $y$,
respectively. Expanding each field operator in a plane-wave form
with respect to the carrier wave number $k$ gives
\begin{equation}
\hat{\psi}_{\rm e}(y) = \frac{1}{\sqrt{{\cal L}}}\sum_k \hat{a}_k\,\texttt{e}^{i k y}\ ,
\:\:\:\:\:\:\:\:\:
\hat{\psi}_{\rm h}(y) = \frac{1}{\sqrt{{\cal L}}}\sum_k \hat{\beta}_{-k}\,\texttt{e}^{-i k y}\ ,
\label{opform}
\end{equation}
and the Fourier transform of Eq.(\ref{rho}) using this result yields
\begin{equation}
\label{rhoq}
\hat{\tilde{\rho}}_{\rm e}(q)
=
\frac{1}{{\cal L}} \sum_{k} \hat{a}^\dagger_{k-q} \, \hat{a}_k\ ,
\:\:\:\:\:\:\:\:\:
\hat{\tilde{\rho}}_{\rm h}(q)
=
\frac{1}{{\cal L}} \sum_{k} \hat{\beta}^\dagger_{-(k+q)} \, \hat{\beta}_{-k}\ .
\end{equation}
\medskip

It is the expectation value of this result that is needed for an induced polarization field in
classical Maxwell's equations.  Solving the SBEs only provides\,\cite{od}
\begin{subequations}
\begin{align}
n^{\rm e}_q(t) & = \left< \hat{a}_q^\dagger \, \hat{a}_q \right> ,
\\
n^{\rm h}_q(t) & = \left< \hat{\beta}_{-q}^\dagger \, \hat{\beta}_{-q} \right> ,
\\
p_{q,q'}(t) & = \left< \hat{\beta}_{-q'} \, \hat{a}_{q} \right> ,
\\
p^\ast_{q,q'}(t) & = \left< \hat{a}^\dag_{q}\,\hat{\beta}^\dag_{-q'} \right> ,
\end{align}
\end{subequations}
where $\hat{a}_q^\dagger$ and $\hat{a}_q$ are the respective electron creation and destruction operators introduced in Eq.\eqref{opform}, and
$\hat{\beta}_{-q}^\dagger$ and $\hat{\beta}_{-q}$ are the respective
hole creation and destruction operators.
However, to calculate the 1D linear density distribution in momentum space we need
the intraband coherence, which is not calculated with the SBEs.
\medskip

In our proposed calculation we do not keep track of the intraband
coherence for conduction electrons or holes. However, we do keep track
of the coherence between a particular electron and all the holes (and vice
versa) through the quantity $p_{q,q'}(t)$. Here, we propose a round-about
way of calculating the expectation value of Eq.\,(\ref{rhoq}) using the
quantities we have from solving the SBEs.
\medskip

The anti-commutator relations for electrons and holes are given by:
\[
\{ \hat{a}_k, \hat{a}_{k'} \}
=
\{ \hat{\beta}_{k}, \hat{\beta}_{k'} \}
=
\{ \hat{a}_{k}, \hat{\beta}^\dagger_{k'} \}
=
\{\hat{a}_{k},\hat{\beta}_{k'} \}
=0\ ,
\]
\begin{equation}
\{ \hat{a}^\dagger_{k}, \hat{a}_{k'} \}
=
\{ \hat{\beta}^\dagger_{k}, \hat{\beta}_{k'} \}
=
\delta_{k,k'}\ .
\end{equation}
Therefore, we can rewrite the operator expression for electrons in Eq.\,(\ref{rhoq}) as
\begin{align}
\hat{a}^\dagger_{k-q} \hat{a}_{k}
=
&
\nonumber \:
\hat{a}^\dagger_{k-q}
\{ \hat{\beta}^\dagger_{-k'},\hat{\beta}_{-k'} \}
\hat{a}_{k}
\\
=
&
\nonumber \:
\hat{a}^\dagger_{k-q} \, \hat{\beta}^\dagger_{-k'} \,
\hat{\beta}_{-k'} \, \hat{a}_{k}
+
\hat{a}^\dagger_{k-q} \, \hat{\beta}_{-k'} \,
\hat{\beta}^\dagger_{-k'} \, \hat{a}_{k}
\\
=
&
\nonumber \:
\hat{a}^\dagger_{k-q} \, \hat{\beta}^\dagger_{-k'} \,
\hat{\beta}_{-k'} \, \hat{a}_{k}
-
\hat{a}^\dagger_{k-q} \, \hat{\beta}_{-k'} \,
\hat{a}_{k} \, \hat{\beta}^\dagger_{-k'}
\\
=
&
\nonumber \:
\hat{a}^\dagger_{k-q} \, \hat{\beta}^\dagger_{-k'} \,
\hat{\beta}_{-k'} \, \hat{a}_{k}
+
\hat{a}^\dagger_{k-q} \, \hat{a}_{k} \,
\hat{\beta}_{-k'} \, \hat{\beta}^\dagger_{-k'}
\\
=
&
\hat{a}^\dagger_{k-q} \, \hat{\beta}^\dagger_{-k'} \,
\hat{\beta}_{-k'} \, \hat{a}_{k}
+
\hat{a}^\dagger_{k-q} \, \hat{a}_{k} \,
\left[1 - \hat{\beta}^\dagger_{-k'} \, \hat{\beta}_{-k'} \right].
\end{align}
This much is exact, but the linear density distribution in momentum space is
calculated by taking the expectation value of these expressions.
By following the approach of
Huag and Koch\,\cite{HaugKoch}, we could use the random-phase approximation to
reduce the expectation values of four operators into products of occupation
numbers and interband coherences:
\begin{align}
\left< \hat{a}^\dagger_{k-q} \hat{a}_{k} \right>
=
&
\nonumber \:
\left< \hat{a}^\dagger_{k-q} \, \hat{\beta}^\dagger_{-k'} \,
\hat{\beta}_{-k'} \, \hat{a}_{k} \right>
+
\left< \hat{a}^\dagger_{k-q} \, \hat{a}_{k} \right>
-
\left< \hat{a}^\dagger_{k-q} \, \hat{a}_{k} \,
\hat{\beta}^\dagger_{-k'} \, \hat{\beta}_{-k'} \right>
\\
\simeq
&
\nonumber \:
\left< \hat{a}^\dagger_{k-q} \, \hat{\beta}^\dagger_{-k'} \right> \,
\left< \hat{\beta}_{-k'} \, \hat{a}_{k} \right>
+
\left< \hat{a}^\dagger_{k-q} \, \hat{a}_{k} \right>
-
\left< \hat{a}^\dagger_{k-q} \, \hat{a}_{k} \right> \,
\left< \hat{\beta}^\dagger_{-k'} \, \hat{\beta}_{-k'} \right>
\\
\simeq
&
\nonumber \:
p^*_{k-q,k'}(t) \; p_{k,k'}(t) +
\left< \hat{a}^\dagger_{k-q} \, \hat{a}_{k} \right>
\left[ 1 - n^{\rm h}_{k'}(t)  \right]\ ,
\\
\left< \hat{a}^\dagger_{k-q} \hat{a}_{k} \right>
\simeq
&
\frac{2}{N_{\rm h}(t)}\,\sum_{k'}\,p^*_{k-q,k'}(t) \; p_{k,k'}(t)\ ,
\end{align}
where $N_{\rm e,h}(t)=2\sum\limits_k\,n^{\rm e,h}_k(t)$.
Therefore, with the random-phase approximation we could calculate
the linear density distribution in momentum space by:
\begin{equation}
\label{rhoeq}
\tilde{\rho}_{\rm e}(q,t)
\simeq
\frac{2}{N_{\rm h}(t){\cal L}}\,
\sum_{k,k'}\,p^*_{k-q,k'}(t) \; p_{k,k'}(t)\ .
\end{equation}
\medskip

An analogous calculation for the holes reveals:
\begin{align}
\left< \hat{\beta}^\dagger_{-(k+q)} \hat{\beta}_{-k} \right>
& \simeq
\:\:
\left[\sum_{k_1}\,n_{\rm e}(k_1+q,\,t)\right]^{-1}\sum_{k'}\,p^\ast_{k'+q,k+q}(t) \; p_{k'+q,k}(t)
\nonumber\\
&
=\frac{2}{N_{\rm e}(t)}\,\sum_{k'}\,p^\ast_{k',k+q}(t) \; p_{k',k}(t)\ ,
\end{align}
and
\begin{equation}
\label{rhohq}
\tilde{\rho}_{\rm h}(q,t)
\simeq
\frac{2}{N_{\rm e}(t){\cal L}}\,
\sum_{k,k'}\,p_{k',k-q}(t) \; p^\ast_{k',k}(t)\ .
\end{equation}

\section{Coulomb-Matrix Elements}
\label{app2}

The Coulomb-interaction matrix elements introduced in Eqs.\,\eqref{eq:DelEk}, \eqref{eq:DelEkp}, \eqref{eq:Rabix} and \eqref{eq:Rabiy} are defined as\,\cite{book-huang}
\begin{subequations}
\begin{align}
	\label{eq:Veh}
	V^{\rm eh}_{k_1,k_1^\prime;\,k_2^\prime,k_2}
	=& \;
	\beta
	\iint d^2\mbox{\boldmath$\xi$}_\perp \, d^2\mbox{\boldmath$\xi$}_\perp^\prime\,
	\frac{
		[\Psi^{\rm e}_{k_1}(\mbox{\boldmath$\xi$}_\perp)]^\ast      [\Psi^{\rm h}_{-k^\prime_1}(\mbox{\boldmath$\xi$}_\perp^\prime)]^\ast
		\Psi^{\rm h}_{-k^\prime_2}(\mbox{\boldmath$\xi$}_\perp^\prime) \, \Psi^{\rm e}_{k_2}(\mbox{\boldmath$\xi$}_\perp)
	}
	{
		|\mbox{\boldmath$\xi$}_\perp-\mbox{\boldmath$\xi$}_\perp^\prime|
	}
	\\
	\label{eq:Vhh}
	V^{\rm hh}_{k^\prime_1,k^\prime_2;\,k^\prime_3,k^\prime_4}
	=& \;
	\beta
	\iint d^2\mbox{\boldmath$\xi$}_\perp \, d^2\mbox{\boldmath$\xi$}_\perp^\prime\,
	\frac{
		[\Psi^{\rm h}_{-k^\prime_1}(\mbox{\boldmath$\xi$}_\perp)]^\ast      [\Psi^{\rm h}_{-k^\prime_2}(\mbox{\boldmath$\xi$}_\perp^\prime)]^\ast
		\Psi^{\rm h}_{-k^\prime_3}(\mbox{\boldmath$\xi$}_\perp^\prime) \, \Psi^{\rm h}_{-k^\prime_4}(\mbox{\boldmath$\xi$}_\perp)}{|\mbox{\boldmath$\xi$}_\perp-\mbox{\boldmath$\xi$}_\perp^\prime|
	}
	\\
	\label{eq:Vee}
	V^{\rm ee}_{k_1,k_2;\,k_3,k_4}
	=& \;
	\beta
	\iint d^2\mbox{\boldmath$\xi$}_\perp \, d^2\mbox{\boldmath$\xi$}_\perp^\prime\,
	\frac{
		[\Psi^{\rm e}_{k_1}(\mbox{\boldmath$\xi$}_\perp)]^\ast      [\Psi^{\rm e}_{k_2}(\mbox{\boldmath$\xi$}_\perp^\prime)]^\ast
		\Psi^{\rm e}_{k_3}(\mbox{\boldmath$\xi$}_\perp^\prime) \, \Psi^{\rm e}_{k_4}(\mbox{\boldmath$\xi$}_\perp)}{|\mbox{\boldmath$\xi$}_\perp-\mbox{\boldmath$\xi$}_\perp^\prime|
	}
\end{align}
\end{subequations}
where $ \beta = e^2 / ( 4 \pi \epsilon_0\epsilon_{\rm r} )$ and $\epsilon_r$ is the average dielectric constant of the host material. Putting in the electron and hole wave functions for the 1D quantum wires gives
\begin{subequations}
\begin{align}
	\label{eq:Veh1D}
	V^{\rm eh}_{k_1,k'_1;\,k'_2,k_2}
&	=
	\delta_{k_1+k'_2,\,k'_1+k_2}\left(\frac{2\beta}{\cal L}\right){\cal Q}_{e,h}(k_1-k_2)\ ,
	\\
	\label{eq:Vhh1D}
	V^{\rm hh}_{k'_1,k'_2;\,k'_3,k'_4}
&	=
	\delta_{k'_1+k'_2,\,k'_3+k'_4}\left(\frac{2\beta}{\cal L}\right){\cal Q}_{h,h}(k'_4-k'_1)\ ,
	\\	
	\label{eq:Vee1D}
	V^{\rm ee}_{k_1,k_2;\,k_3,k_4}
&	=
	\delta_{k_1+k_2,\,k_3+k_4} \left(\frac{2\beta}{\cal L}\right){\cal Q}_{e,e}(k_1-k_4)\ ,
\end{align}
\end{subequations}
where $\mbox{\boldmath$\xi$}=(\mbox{\boldmath$\xi$}_\perp,\xi_\|)$ is a local position vector for quantum wires,
${\cal Q}_{\mu,\nu}(x)=	\iint  d^2\mbox{\boldmath$\xi$}_\perp   d^2\mbox{\boldmath$\xi$}'_\perp  \,
\left|\psi^{\mu}_0(\mbox{\boldmath$\xi$}_\perp)\right|^2K_0(|x|[|\mbox{\boldmath$\xi$}_\perp - \mbox{\boldmath$\xi$}'_\perp|^2+\delta_0^2]^{1/2})
\left|\psi^{\nu}_0(\mbox{\boldmath$\xi$}'_\perp)\right|^2$ is an interaction integral
for $\mu,\nu={\rm e,h}$, $\delta_0$ is the thickness of the wire,
$K_0(|q||x|)$ is the modified Bessel function of the third kind,
and the cutoff for the modified Bessel function
is $|q^{\rm e,h}_{\rm min}|\sim \alpha_{\rm e,h}/2$. In our calculations, the screening effects on the Coulomb interactions in Eqs.\,\eqref{eq:Veh1D}-\eqref{eq:Vee1D} have been taken into account by employing the dielectric function in Eq.\,\eqref{eq:E1D}
under the random-phase approximation.

\section{Carrier- and Pair-Scattering Rates}
\label{app-3}

For photo-excitations near a bandgap, the microscopic scattering-in and scattering-out rates for the electrons and holes are calculated as\,\cite{huang-6}
\begin{align}
W^{\rm e, (in)}_{j,k}(t)
\nonumber
& =
\frac{2\pi}{\hbar}
\sum_{k_1}{}^\prime   \,   \left|V^{\rm ep}_{k,k_1}\right|^2    \,   n^{\rm e}_{j, k_1}(t)
\left\{
N_0(\Omega_{\rm ph}) \,
L(\varepsilon^{\rm e}_k - \varepsilon^{\rm e}_{k_1} - \hbar\Omega_{\rm ph}, \, \hbar\Gamma_{\rm ph})
\right.
\\ & \nonumber
\left.
+\left[
N_0(\Omega_{\rm ph})+1
\right]
L(\varepsilon^{\rm e}_k - \varepsilon^{\rm e}_{k_1} + \hbar\Omega_{\rm ph}, \, \hbar\Gamma_{\rm ph})
\right\}
\\ & \nonumber
+\frac{2\pi}{\hbar}  \sum_{k_1}{}^\prime   \sum_{k^\prime,k_1^\prime}{}^\prime  \,
\left|V^{\rm eh}_{k,k^\prime;\,k_1^\prime,k_1}\right|^2
\left[
1 - n^{\rm h}_{j, k^\prime}(t)
\right]
n^{\rm h}_{j, k_1^\prime}(t)   \,   n^{\rm e}_{j, k_1}(t)
\\ & \nonumber
\times
L(\varepsilon^{\rm e}_k+\varepsilon^{\rm h}_{k^\prime}-\varepsilon^{\rm e}_{k_1}-\varepsilon^{\rm h}_{k_1^\prime},
\hbar \gamma_{\rm eh})
\\ & \nonumber
+\frac{2\pi}{\hbar}   \sum_{k_2,k_3,k_4}{}^\prime  \,
\left|V^{\rm ee}_{k,k_2;\,k_3,k_4}\right|^2
\left[
1-n^{\rm e}_{j, k_2}(t)
\right]
n^{\rm e}_{j, k_3}(t)   \,   n^{\rm e}_{j, k_4}(t)
\\ &
\label{eq:Wine}
\times
L(\varepsilon^{\rm e}_k+\varepsilon^{\rm e}_{k_2}-\varepsilon^{\rm e}_{k_3}-\varepsilon^{\rm e}_{k_4},
\hbar \gamma_{\rm e})\ ,
\end{align}
\begin{align}
W^{\rm e, (out)}_{j, k}(t)
\nonumber
& =
\frac{2\pi}{\hbar}   \sum_{k_1}{}^\prime   \,   \left|V^{\rm ep}_{k,k_1}\right|^2
\left[
1-n^{\rm e}_{j, k_1}(t)
\right]
\left\{ N_0(\Omega_{\rm ph})
L(\varepsilon^{\rm e}_{k_1}-\varepsilon^{\rm e}_k-\hbar\Omega_{\rm ph} \; ,
\hbar  \Gamma_{\rm ph})
\right.
\\ & \nonumber
\left.
+\left[
N_0(\Omega_{\rm ph})+1
\right]
L(\varepsilon^{\rm e}_{k_1} - \varepsilon^{\rm e}_k + \hbar \Omega_{\rm ph} \; , \hbar \Gamma_{\rm ph})
\right\}
\\ & \nonumber
+  \frac{2\pi}{\hbar}   \sum_{k_1}{}^\prime   \sum_{k^\prime,k_1^\prime}{}^\prime   \,
\left|
V^{\rm eh}_{k_1,k^\prime;\,k_1^\prime,k}
\right|^2
\left[
1-n^{\rm h}_{j, k^\prime}(t)
\right]
n^{\rm h}_{j, k_1^\prime}(t)
\left[
1-n^{\rm e}_{j, k_1}(t)
\right]
\\ & \nonumber
\times
L(\varepsilon^{\rm e}_{k_1}+\varepsilon^{\rm h}_{k^\prime}-\varepsilon^{\rm e}_k-\varepsilon^{\rm h}_{k_1^\prime} \; ,
\hbar  \gamma_{\rm eh})
\\ & \nonumber
+\frac{2\pi}{\hbar}   \sum_{k_2,k_3,k_4}{}^\prime   \,
\left|
V^{\rm ee}_{k_4,k_2;\,k_3,k}
\right|^2
\left[
1-n^{\rm e}_{j, k_2}(t)
\right]
n^{\rm e}_{j, k_3}(t)
\left[
1-n^{\rm e}_{j, k_4}(t)
\right]
\\ &
\label{eq:Woute}
\times
L(\varepsilon^{\rm e}_{k_4}+\varepsilon^{\rm e}_{k_2}-\varepsilon^{\rm e}_{k_3}-\varepsilon^{\rm e}_k \; ,
\hbar  \gamma_{\rm e})\ ,
\end{align}
\begin{align}
W^{\rm h, (in)}_{j,k'}(t)
\nonumber
& =
\frac{2\pi}{\hbar}
\sum_{k^\prime_1}{}^\prime   \,   \left|V^{\rm hp}_{k',k^\prime_1}\right|^2    \,   n^{\rm h}_{j, k^\prime_1}(t)
\left\{
N_0(\Omega_{\rm ph}) \,
L(\varepsilon^{\rm h}_{k'} - \varepsilon^{\rm h}_{k^\prime_1} - \hbar\Omega_{\rm ph} \; , \, \hbar\Gamma_{\rm ph})
\right.
\\ & \nonumber
\left.
+\left[
N_0(\Omega_{\rm ph})+1
\right]
L(\varepsilon^{\rm h}_{k'} - \varepsilon^{\rm h}_{k^\prime_1} + \hbar\Omega_{\rm ph} \; , \, \hbar\Gamma_{\rm ph})
\right\}
\\ & \nonumber
+\frac{2\pi}{\hbar}  \sum_{k^\prime_1}{}^\prime   \sum_{k, k_1}{}^\prime  \,
\left|V^{\rm eh}_{k,k^\prime;\,k_1^\prime,k_1}\right|^2
\left[
1 - n^{\rm e}_{j, k}(t)
\right]
n^{\rm e}_{j, k_1}(t)   \,   n^{\rm h}_{j, k^\prime_1}(t)
\\ & \nonumber
\times
L(\varepsilon^{\rm e}_k+\varepsilon^{\rm h}_{k^\prime}-\varepsilon^{\rm e}_{k_1}-\varepsilon^{\rm h}_{k_1^\prime} \; ,
\hbar \gamma_{\rm eh})
\\ & \nonumber
+\frac{2\pi}{\hbar}   \sum_{k^\prime_2,  k^\prime_3,  k^\prime_4}{}^\prime  \,
\left|V^{\rm hh}_{k^\prime,  k^\prime_2;\,  k^\prime_3,  k^\prime_4 }\right|^2
\left[
1-n^{\rm h}_{j, k^\prime_2}(t)
\right]
n^{\rm h}_{j, k^\prime_3}(t)   \,   n^{\rm h}_{j, k^\prime_4}(t)
\\ &
\label{eq:Winh}
\times
L(\varepsilon^{\rm h}_{k^\prime} + \varepsilon^{\rm h}_{k^\prime_2} - \varepsilon^{\rm h}_{k^\prime_3} - \varepsilon^{\rm h}_{k^\prime_4} \; ,
\hbar \gamma_{\rm h})\ ,
\end{align}
and
\begin{align}
W^{\rm h, (out)}_{j, k^\prime}(t)
& =
\nonumber
\frac{2\pi}{\hbar}   \sum_{k^\prime_1}{}^\prime
\left|V^{\rm hp}_{k^\prime,k^\prime_1}\right|^2
\left[
1-n^{\rm h}_{j, k^\prime_1}(t)
\right]
\left\{ N_0(\Omega_{\rm ph})
L(\varepsilon^{\rm h}_{k^\prime_1} - \varepsilon^{\rm h}_{k^\prime} - \hbar\Omega_{\rm ph} \; ,
\hbar  \Gamma_{\rm ph})
\right.
\\ & \nonumber
\left.
+\left[
N_0(\Omega_{\rm ph})+1
\right]
L(\varepsilon^{\rm h}_{k^\prime_1} - \varepsilon^{\rm h}_{k^\prime}
+ \hbar \Omega_{\rm ph} \; , \hbar \Gamma_{\rm ph})
\right\}
\\ & \nonumber
+  \frac{2\pi}{\hbar}   \sum_{k^\prime_1}{}^\prime   \sum_{k, k_1}{}^\prime   \,
\left|
V^{\rm eh}_{k,k_1^\prime;\,k^\prime,k_1}
\right|^2
\left[
1-n^{\rm e}_{j, k}(t)
\right]
n^{\rm e}_{j, k_1}(t)
\\ & \nonumber
\times
\left[
1-n^{\rm h}_{j, k^\prime_1}(t)
\right]
L(\varepsilon^{\rm e}_{k}+\varepsilon^{\rm h}_{k_1^\prime}-\varepsilon^{\rm e}_{k_1}-\varepsilon^{\rm h}_{k^\prime},
\hbar  \gamma_{\rm eh})
\\ & \nonumber
+\frac{2\pi}{\hbar}   \sum_{k^\prime_2, k^\prime_3, k^\prime_4}{}^\prime   \,
\left|
V^{\rm hh}_{k^\prime_4, k^\prime_2; \, k^\prime_3, k^\prime}
\right|^2
\left[
1-n^{\rm h}_{j, k^\prime_2}(t)
\right]
n^{\rm h}_{j, k^\prime_3}(t)
\\ &
\label{eq:Wouth}
\times
\left[
1-n^{\rm h}_{j, k^\prime_4}(t)
\right]
L(\varepsilon^{\rm h}_{k^\prime_4}+\varepsilon^{\rm h}_{k^\prime_2}-\varepsilon^{\rm h}_{k^\prime_3}
-\varepsilon^{\rm h}_{k^\prime} \; , \hbar  \gamma_{\rm h})\ ,
\end{align}
where the impact-ionization, Auger and exciton-pair scattering, which are important only for narrow-bandgap semiconductors, have been neglected.
Here, $L(a,b) = (b/\pi) / (a^2 + b^2)$ is the Lorentzian function, the primed summations exclude the terms satisfying either $k^\prime=k_1^\prime$ or $k_1=k$,
as well as the terms satisfying $k_2=k_3$, $k_4=k$, $k^\prime_2=k^\prime_3$ or $k^\prime_4=k^\prime$,
$N_0(\Omega_{\rm ph})=\left[\exp(\hbar\Omega_{\rm ph}/k_{\rm B}T)-1\right]^{-1}$ is the Bose function for the thermal-equilibrium longitudinal-optical phonons,
$\Omega_{\rm ph}$ and $1/\Gamma_{\rm ph}$ are the frequency and lifetime of longitudinal-optical phonons in the host semiconductors,
$1/\gamma_{\rm e}$ and $1/\gamma_{\rm h}$ are the lifetimes of photo-excited electrons and holes, respectively, and
$\gamma_{\rm eh}=(\gamma_{\rm e}+\gamma_{\rm h})/2$.
In addition, both the interband (second terms) and the intraband (third terms) energy relaxations are included.
\medskip

The RPA screened coupling between the longitudinal-optical phonons and electrons or holes in Eqs.\,\eqref{eq:Wine}-\eqref{eq:Wouth} are\,\cite{book-huang}
\begin{subequations}
	\begin{align}
	\label{eq:Vhp}
		\left|V^{\rm hp}_{k',k^\prime_1}\right|^2
&	=
	\frac{e^2\hbar\Omega_{\rm ph}}{2\pi\epsilon_0{\cal L}}
	\left(
	\frac{1}{\epsilon_{\infty}}  -  \frac{1}{\epsilon_{\rm s}}\right)\frac{|{\cal Q}_{\rm h,h}(k'_1-k')|}{[\epsilon_{\rm 1D}(|k'_1-k'|,t)]^2}\ ,
   \\
		\label{eq:Vep}	
	\left|V^{\rm ep}_{k,k_1}\right|^2
&	=
	\frac{e^2\hbar\Omega_{\rm ph}}{2\pi\epsilon_0{\cal L}}
	\left(
	\frac{1}{\epsilon_{\infty}}  -  \frac{1}{\epsilon_{\rm s}}
	\right)\frac{|{\cal Q}_{\rm e,e}(k_1-k)|}{[\epsilon_{\rm 1D}(|k_1-k|,t)]^2}\ ,
	\end{align}
\end{subequations}
where $\epsilon_{\infty}$ and $\epsilon_{\rm s}$ are the high-frequency and static dielectric constants of the host polar semiconductor.

\section{Diagonal and Off-Diagonal Dephase Rates}
\label{app-4}

The diagonal dephasing of $p^\sigma_{j,k,k^\prime}(t)$ in Eq.\,\eqref{eq:dpdt} has been taken into account by $\Delta_{j,k}^{\rm e}(t)$ and $\Delta_{j,k'}^{\rm h}(t)$ terms
with $\Delta_{j,k,k'}^{\rm eh}(t)=\Delta_{j,k}^{\rm e}(t)+\Delta_{j,k'}^{\rm h}(t)$, which are given by\,\cite{od}
\[
\Delta_{j,k}^{\rm e}(t)=\frac{\pi}{\hbar}\sum_{k_1,q\neq 0}\left|V^{\rm ee}_{k_1-q,k+q;\,k,k_1}\right|^2\left[
	L(\varepsilon^{\rm e}_{k_1-q}+\varepsilon^{\rm e}_{k+q}-\varepsilon^{\rm e}_k-\varepsilon^{\rm e}_{k_1},\gamma_{\rm e})\right.
\]
\[
\left.\times\left\{n_{j,k_1-q}^{\rm e}(t)\,n_{j,k+q}^{\rm e}(t)\,[1-n_{j,k_1}^{\rm e}(t)]
	+[1-n_{j,k_1-q}^{\rm e}(t)]\,[1-n_{j,k+q}^{\rm e}(t)]\,n_{j,k_1}^{\rm e}(t)\right\}\right]
\]
\[
+\frac{\pi}{\hbar}\sum_{k'_1,q\neq 0}\left|V^{\rm eh}_{k-q,k_1'-q;\,k_1',k}\right|^2\left[
	L(\varepsilon^{\rm h}_{k'_1-q}+\varepsilon^{\rm e}_{k-q}-\varepsilon^{\rm e}_{k}-\varepsilon^{\rm h}_{k'_1},\gamma_{\rm eh})
	\left\{n_{j,k'_1-q}^{\rm h}(t)\,[1-n_{j,k'_1}^{\rm h}(t)]\,n_{j,k-q}^{\rm e}(t)\right.\right.
\]
\begin{equation}
\left.\left.+[1-n_{j,k'_1-q}^{\rm h}(t)]\,n_{j,k'_1}^{\rm h}(t)\,[1-n_{j,k-q}^{\rm e}(t)]\right\}\right]\ ,
\label{diage}
\end{equation}
\[
\Delta_{j,k'}^{\rm h}(t)=\frac{\pi}{\hbar}\sum_{k'_1,q'\neq 0}\left|V^{\rm hh}_{k'_1-q',k'+q';\,k',k'_1}\right|^2\left[
	L(\varepsilon^{\rm h}_{k'_1-q'}+\varepsilon^{\rm h}_{k'+q'}-\varepsilon^{\rm h}_{k'}-\varepsilon^{\rm h}_{k'_1},\gamma_{\rm h})\right.
\]
\[
\left.\times\left\{n_{j,k'_1-q'}^{\rm h}(t)\,n_{j,k'+q'}^{\rm h}(t)[1-n_{j,k'_1}^{\rm h}(t)]
	+[1-n_{j,k'_1-q'}^{\rm h}(t)]\,[1-n_{j,k'+q'}^{\rm h}(t)]\,n_{j,k'_1}^{\rm h}(t)\right\}\right]
\]
\[
+\frac{\pi}{\hbar}\sum_{k_1,q'\neq 0}\left|V^{\rm eh}_{k_1-q',k'-q';\,k',k_1}\right|^2\left[
	L(\varepsilon^{\rm e}_{k_1-q'}+\varepsilon^{\rm h}_{k'-q'}-\varepsilon^{\rm h}_{k'}-\varepsilon^{\rm e}_{k_1},\gamma_{\rm eh})
	\left\{n_{j,k_1-q'}^{\rm e}(t)\,[1-n_{j,k_1}^{\rm e}(t)]\right.\right.
\]
\begin{equation}
\left.\left.\times n_{j,k'-q'}^{\rm h}(t)+[1-n_{j,k_1-q'}^{\rm e}(t)]\,n_{j,k_1}^{\rm e}(t)\,[1-n_{j,k'-q'}^{\rm h}(t)]\right\}\right]\ .
\label{diagh}
\end{equation}
Furthermore, the off-diagonal dephasing of $p^\sigma_{j,k,k^\prime}(t)$ in Eq.\,\eqref{eq:dpdt} has also been included by
$\Lambda_{j,k,q}^{\rm e}(t)$ and $\Lambda_{j,k',q'}^{\rm h}(t)$ terms, which are given by\,\cite{od}
\[
\Lambda_{j,k,q}^{\rm e}(t)=\frac{\pi}{\hbar}\sum_{k_1}\left|V^{\rm ee}_{k_1,k+q;\,k,k_1+q}\right|^2\left[
	L(\varepsilon^{\rm e}_{k_1+q}+\varepsilon^{\rm e}_k-\varepsilon^{\rm e}_{k_1}-\varepsilon^{\rm e}_{k+q},\gamma_{\rm e})\right.
\]
\[
\left.\times\left\{n_{j,k_1+q}^{\rm e}(t)\,n_{j,k}^{\rm e}(t)\,[1-n_{j,k_1}^{\rm e}(t)]
	+[1-n_{j,k_1+q}^{\rm e}(t)]\,[1-n_{j,k}^{\rm e}(t)]\,n_{j,k_1}^{\rm e}(t)\right\}\right]
\]
\[
+\frac{\pi}{\hbar}\sum_{k'_1}\left|V^{\rm eh}_{k,k_1'-q;\,k_1',k+q}\right|^2\left[
	L(\varepsilon^{\rm h}_{k'_1-q}+\varepsilon^{\rm e}_{k}-\varepsilon^{\rm h}_{k'_1}-\varepsilon^{\rm e}_{k+q},\gamma_{\rm eh})
	\left\{n_{j,k'_1-q}^{\rm h}(t)\,[1-n_{j,k'_1}^{\rm h}(t)]\,n_{j,k}^{\rm e}(t)\right.\right.
\]
\begin{equation}
\left.\left.+[1-n_{j,k'_1-q}^{\rm h}(t)]\,n_{j,k'_1}^{\rm h}(t)\,[1-n_{j,k}^{\rm e}(t)]\right\}\right]\ ,
\end{equation}
\[
\Lambda_{j,k',q'}^{\rm h}(t)=\frac{\pi}{\hbar}\sum_{k'_1}\left|V^{\rm hh}_{k'_1,k'+q';\,k',k'_1+q'}\right|^2\left[
	L(\varepsilon^{\rm h}_{k'_1+q'}+\varepsilon^{\rm h}_{k'}-\varepsilon^{\rm h}_{k'_1}-\varepsilon^{\rm h}_{k'+q'},\gamma_{\rm h})\right.
\]
\[
\left.\times\left\{n_{j,k'_1+q'}^{\rm h}(t)\,n_{j,k'}^{\rm h}(t)[1-n_{j,k'_1}^{\rm h}(t)]
	+[1-n_{j,k'_1+q'}^{\rm h}(t)]\,[1-n_{j,k'}^{\rm h}(t)]\,n_{j,k'_1}^{\rm h}(t)\right\}\right]
\]
\[
+\frac{\pi}{\hbar}\sum_{k_1}\left|V^{\rm eh}_{k_1,k'+q';\,k',k_1-q'}\right|^2\left[
	L(\varepsilon^{\rm e}_{k_1-q'}+\varepsilon^{\rm h}_{k'}-\varepsilon^{\rm e}_{k_1}-\varepsilon^{\rm h}_{k'+q'},\gamma_{\rm eh})
	\left\{n_{j,k_1-q'}^{\rm e}(t)\,[1-n_{j,k_1}^{\rm e}(t)]\right.\right.
\]
\begin{equation}
\left.\left.\times n_{j,k'}^{\rm h}(t)+[1-n_{j,k_1-q'}^{\rm e}(t)]\,n_{j,k_1}^{\rm e}(t)\,[1-n_{j,k'}^{\rm h}(t)]\right\}\right]\ .
\end{equation}

\end{document}